\documentclass[%
aps, prfluids, preprint, 
showkeys,
notitlepage,
superscriptaddress,
nofootinbib,
]{revtex4-2}

\usepackage{graphicx}
\usepackage{dcolumn}
\usepackage{bm}
\usepackage{pgfplots}
\usepackage{pgfplotstable}
\usepackage{tikz}
\pgfplotsset{compat=newest}
\pgfplotsset{compat=1.18}
\usepackage{amsmath}
\usepackage{amssymb}
\usepackage{subcaption}
\usepackage{multirow}

\usetikzlibrary{pgfplots.groupplots}
\usepgfplotslibrary{fillbetween}

\newcommand{\vect}[1]{\bm{#1}}
\newcommand{\dif}{\mathrm{d}}

\def \vx{\boldsymbol{x}}

\def \vv{\vect{V}}

\begin{document}


\title{Physically Organized Latent Spaces in Unsupervised Autoencoders: \\Evidence from Aerodynamic Databases}

\author{Renato Tognaccini\footnote{Corresponding author, renato.tognaccini@unina.it}}
\author{Ettore Saetta}
 \affiliation{Dipartimento di Ingegneria Industriale \\ Università di Napoli Federico II}
\author{Gianluca Iaccarino }%
\affiliation{Department of Mechanical Engineering \\ Stanford University}%


\date{\today}

\begin{abstract}
Using aerodynamic databases of progressively increasing complexity —  from linear analytical theory to nonlinear separated RANS flows — we demonstrate that an autoencoder, trained solely by reconstruction, can spontaneously organize the latent representation according to physically meaningful variables and aerodynamic laws. 
Although the latent coordinates themselves vary under random initialization, the underlying physical organization is reproducible, even quantitatively. 
In addition, we present a three-dimensional latent space representation of the airfoil flow obtained by an unsupervised autoencoder with sequential training in which the inviscid field is decoupled from the boundary layer effects with the third latent variable strongly correlated with the Reynolds number and aerodynamic drag.
Cross-validated regression and affine-alignment measures show that physical parameters and responses are strongly and linearly encoded in the learned coordinates, consistently across independent trainings.
\end{abstract}


\maketitle


\newpage

\section{Introduction}

Deep neural networks (DNN) and Machine Learning (ML) are now widely diffused  as  new tools for scientific analysis. 
High-dimensional physical fields often evolve on lower-dimensional manifolds, a result traditionally obtained via human induction. Principal component analysis (PCA) and proper orthogonal decomposition (POD) are typical mathematical tools introduced to compress complex data representations; they have their nonlinear data-driven counterparts in the autoencoders (AE) and  are all effectively adopted in  reduced-order analyses  providing 
 a low-dimensional representation  of the data usually named \emph{reduced-order coordinate space} in POD and  \emph{latent space} in case of AE. 

It would be an important step forward to use the DNNs as tools to study and understand unclear physical phenomena. Data driven analyses for physical discovery is well discussed by Iten et al. \cite{iten2020}. The authors address the problem showing how a DNN could rediscover classical physical laws, like damped pendulum, conservation of angular momentum and  the heliocentric model of solar system.
Their conclusion is that DNNs should be exploited as powerful new tools for scientific analysis. However they 
emphasize the problem of the
 requirement of prior
knowledge on the systems of interest: ``\emph{In certain situations one might not have such prior knowledge or does not want to impose it to allow the machine
to find entirely different representations of the physical
system.}''


The subject of \emph{disentanglement} via DNN has been extensively discussed, see for instance \cite{Higgins2017,kim18b, mathieu19a,khemakhem20a}. 
A latent representation is disentangled when individual latent variables, or identifiable subsets of them, correspond to distinct underlying factors of variation in the data (for instance Mach number or angle of attack in Aerodynamics), such that changes in one factor are represented with minimal influence on the others.
However,
unsupervised representation learning does not, in general, guarantee recovery of the underlying generative factors. In particular, Locatello et al. \cite{locatello19a} demonstrated that unsupervised disentanglement is fundamentally non-identifiable in the absence of suitable inductive biases on the learning method and the data. Consequently, the emergence of reproducible physically organized latent representations from reconstruction-only training cannot be assumed a priori and must be assessed empirically.
Given the intrinsic non-uniqueness of unsupervised latent representations, which physical properties of the representation remain invariant or reproducible across independent trainings?

Clearly, the interest in this subject spreads beyond Fluid Dynamics. For instance, Lu et al. \cite{lu2020} already adopted an unsupervised physics-informed variational AE (VAE) to 
 identify relevant
parameters and extract them from raw and  noisy  data. The tool was then applied to nonlinear fiber propagation data analysis.
Kalinin et al. \cite{kalinin2022} used a VAE on ferroelectric-domain images to find that individual latent directions acquire recognizable meanings.
Yin et al. \cite{Yin2021} again by a VAE on alloy configurations  propose a quantity from the latent representation that behaves as an order parameter for order–disorder phase transitions. They  called the result a physically interpretable VAE-based order parameter.
Baima et al. \cite{baima2022}  examined both the capabilities and limitations of AEs in discovering free energy barriers at finite temperature for defect processes in materials.
Nautrup et al. \cite{nautrup2022} evidenced that DNN
often lack \emph{interpretability}; in particular noted absence of 
an \emph{operational meaning} in disentangling representations: disentanglement and interpretability are the two keywords in literature today.

Focusing on Fluid Dynamics, Jacobsen et al. \cite{jacobsen2022}, while studying model porous flow problems, evidenced difficulties with consistently disentangling generative parameters  by an unsupervised VAE framework and suggested  a semi-supervised approach in alternative.
In Aerodynamics, Kang et al. \cite{kang2022}, showed how a $\beta$-VAE could be able to recognize Mach number and angle of attack (AoA) as generative parameters in a database of airfoil flows.
Bousquet et al. \cite{bousquet2025} studied  AEs and VAEs latent spaces applied to an oscillating cylinder flow emphasizing the  need to align latent space representations with physical interpretation.
A recent work has also addressed the interpretability of nonlinear autoencoder representations; Magri and Doan \cite{magri2022} interpreted the latent space as a curved manifold approximating the system attractor and developed a decomposition to extract its dominant latent directions.
Racca et al. \cite{Racca2023}  used a convolutional AE to obtain the manifold on which turbulent dynamics evolve and then use an echo-state network to evolve the system in that latent space. 
Giral et al. \cite{aerojepa2026} showed that self-supervised predictive learning can generate semantically structured latent representations of very complex three-dimensional aerodynamic databases, in which information related to geometry and aerodynamic quantities emerges without direct supervision.
Page et al. \cite{page2021} by a deep convolutional AE discovered a low-dimensional representation of turbulent Kolmogorov flow.
Marra et al. \cite{Marra2024}, using ISOMAP manifold learning, showed that low-dimensional manifolds obtained from flow snapshots can exhibit coordinates associated with distinct physical mechanisms, including vortex shedding, boat-tailing, Magnus effect and stagnation-point control.
Fukami et al. \cite{Fukami2023,Fukami2024,Fukami2025} studied an  airfoil interacting with extreme vortex gusts and in transonic buffet conditions
using an observable-augmented nonlinear AE. They explicitly
stated that a regular autoencoder does not necessarily produce a physically understandable latent distribution.
In a different paper, Fukami \& Taira \cite{Fukami2025b} applied the method to the study of turbulent flows.
Nair et al. \cite{nair2025} compressed 93 different rough surfaces into only three latent variables and related the latent representation to equivalent sand-grain roughness.
Vinuesa et al. \cite{vinuesa2026} finally address the relevance of what they call the future \emph{explainable AI} as a very powerful tool able to outperform human resources.

The literature has established that physically meaningful latent representations can be obtained when suitable regularization, supervision, disentanglement objectives, or physics-based inductive biases are introduced. However, much less is understood about whether—and under which conditions—physical structure emerges spontaneously in the latent representation learned by an AE for example, whether this structure is reproducible across independent trainings, and whether the relationship between latent coordinates and physical quantities can be not only qualitative but quantitative.
Most previous work asks how to design latent spaces to become physically interpretable; our question is whether physical coordinates can emerge spontaneously from reconstruction alone, and whether that emergence survives independent random initialization.

Our interest in this subject came out when, in previous works  dedicated to the applications of AEs in Aerodynamics \cite{saetta2022,saettaUQ}, we realized the presence, in very-highly compressed latent spaces,
of patterns looking more meaningful than the simple parameter disentanglement;  we named it \emph{physical imprinting} of the latent space. We therefore decided to deepen the investigation with the help of very simple ground truths.

The approach adopted here is to compare physical descriptions provided by aerodynamic theories of progressively increasing complexity with the latent-space representations obtained from unsupervised vanilla AEs with random weight initialization. 
For this purpose we shall introduce the concept of physical latent space (PLS) which is defined as
a parameterization constructed from known physical parameters or observables
 for which a human-derived mapping is established between latent coordinates or directions and identifiable physical quantities, mechanisms, or known physical regimes. It will be  the human \emph{counterpart} of the data-driven latent space (DDLS) given by the latent representation learned by the AE. 
The objective is then to assess whether, and to what extent, the organization of the DDLS reflects that of the corresponding PLS.

We shall distinguish here between  \emph{physical parameter},
an externally prescribed or defining quantity used to specify the system or operating condition (in our cases AoA, airfoil camber, airfoil thickness, Reynolds number) and \emph{physical response},
a quantity  emerging from the solution of the governing equations  and characterizing the resulting physical state or behavior, 
e.g. lift, drag, minimum pressure or a boundary-layer correction.
Kang et al. showed that their $\beta$-VAE was able to identify in the latent space the physical parameters of their database, but we could not find in literature  the evidence of  physical responses in spontaneous latent spaces.

Following a brief description of the adopted AEs, we shall analyze three aerodynamic theories of increasing complexity: a) inviscid incompressible linearized theory applied to an infinitely thin airfoil; b) inviscid non-linearized theory (Joukowski airfoil); c) viscous  flow around airfoils with massive separations.

Finally, we shall show how a three-dimensional latent space obtained by the application of an AE with sequential training and adaptive resolution (AESTAR) recently introduced by Saetta et al. \cite{sae2026} can decompose
the inviscid physics in a latent space plane while the viscosity correction is controlled by the third  latent variable strictly correlated with  physical output variables, the drag coefficient of the airfoil and the Reynolds number.

Despite the non-uniqueness of the latent coordinate system, independently trained unconstrained autoencoders recover reproducible physical organization of the aerodynamic data manifold and
the learned latent representation admits a quantitative mapping to physical parameters or observables.

The practical value of latent-space interpretability is not limited to explaining the internal representation of the network. Once physically distinct effects are associated with different latent directions, these effects can be manipulated independently through the decoder.
That gives a reason why interpretability matters.




\section{Vanilla autoencoder and autoencoder with sequential training}
\label{sec.AE}

The  adopted AE technology   is extensively described in \cite{saetta2022,saettaUQ,sae2026}. It is here very shortly recalled. A Vanilla AE consists of two DNNs, the encoder and the decoder. The algorithm is unsupervised: the encoder takes the input data and constructs (\textit{learns}) an equivalent low-dimensional representation, i.e. a vector in the \textit{latent space}; the decoder starts from the latent vector and reconstructs the input.
\begin{figure}
    \centering
    \includegraphics[width=0.8\linewidth]{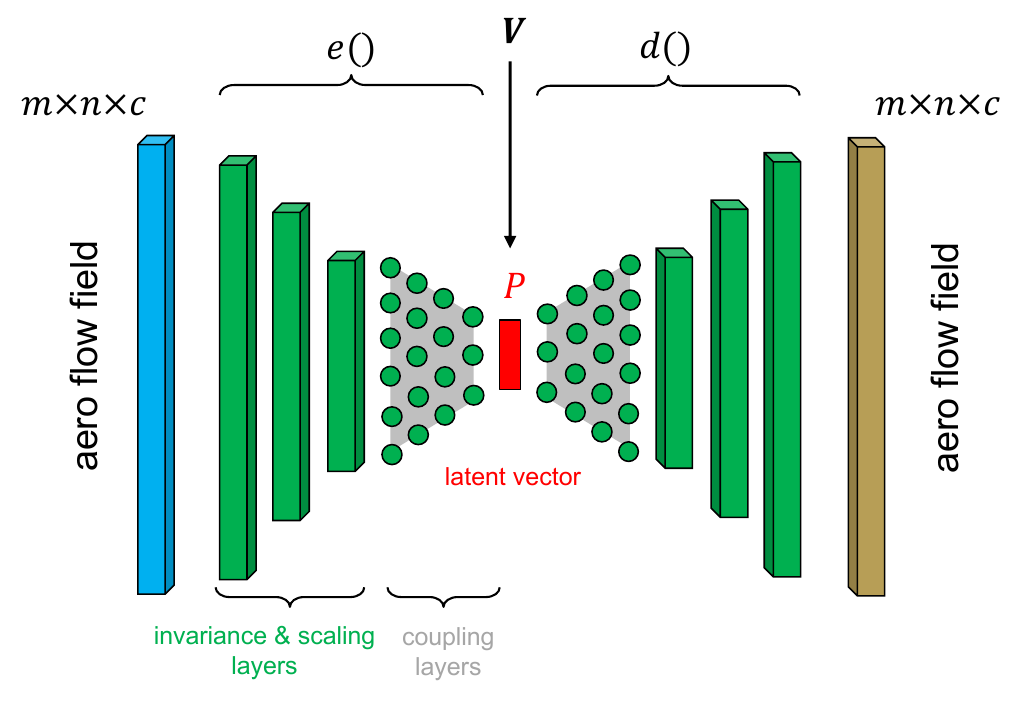}
    \caption{Illustrative scheme of the vanilla AE.}
    \label{fig.ae_sche}
\end{figure}
It is mathematically summarized as:
\begin{equation}
    \vv = e\left(\vx\right); \quad 
    \hat{\vx} = d\left(\vv\right); \quad 
    \hat{\vx} = f\left(\vx\right) =  d\left(e\left(\vx\right)\right)  \ ,
    \label{eq:code}
\end{equation}
where $e\left(\right)$ is the \textit{encoder} function, $d\left(\right)$ is the \textit{decoder} function, $f\left(\right)$ is the entire transformation \textit{encoder}+\textit{decoder}, $\vv$ is the vector of latent variables (also referred to as \textit{latent space}) and $\vx$ and $\hat{\vx}$ are respectively the input and its reconstruction:
\begin{equation*}
    \vx,\ \hat{\vx} \in \mathbb{R}^{m\times n\times c}; \qquad 
    \vv \in \mathbb{R}^{P}  \ ,
\end{equation*}
where $m\times n\times c$ is the dimension of the input data and $P$ is the dimension of the \textit{latent space}.
The scheme depicted in figure \ref{fig.ae_sche} is an illustrative example of a convolutional vanilla AE adopted in the present article.
A Rectified Linear Unit (\textit{ReLU}) activation function is applied after each layer, except for the output layer, where an \textit{Identity} function is adopted since, here, the focus is on regression problems.
In all the performed tests, a regularization term is applied during the training, penalizing the $L_2$ norm of the encoder weights. This has proven to lead to better convergence properties in the learning phase and increased latent space compression for our cases of interest \cite{saetta2022,Newson2020}.

We shall also discuss some results obtained by a multi-fidelity AE with sequential training AESTAR proposed by Saetta et al. \cite{sae2026}. It incorporates a \textit{conditioning} procedure distributed across different training phases: successive training stages are performed, and each of them is \textit{informed} by the previous ones.
Latent space \textit{conditioning} is achieved by introducing an additional term in the loss function, designed to compel the latent variables to satisfy desired constraints. In other words, it guides the latent space to assume a specific topological structure. 
The conditioning in the AESTAR method only involves latent subspaces of the current and previous training instances.
Consequently, the algorithm is an \emph{unsupervised} latent space conditioning technique.
The {\it sequentiality} of the training can be applied either to the latent space dimension or to the training database.
In particular, we define two procedures for employing the present training algorithm:
\begin{enumerate}
    \item sequential training by extracting hierarchical latent subspaces (with a fixed training dataset), denoted as \emph{sequential conditioning AE};
    \item adaptive resolution of multi-fidelity database, by extracting features representing the different levels of data resolution, denoted as \emph{multi-fidelity conditioning AE}.
\end{enumerate}

The AESTAR is naturally suited for multi-fidelity  datasets (higher resolution or more complex physical model).  Specifically, each successive phase of the training  can rely on a dataset with  increasingly higher fidelity data. The resulting algorithm provides adaptive resolution, in the sense that the same trained autoencoder can even return reconstructions  selectively focusing on a specific fidelity. Furthermore, the sequential increase of the latent space dimension introduces information that can be directly attributed to the increased fidelity of the dataset.

\begin{figure}
\centering
\includegraphics[width=0.8\linewidth]{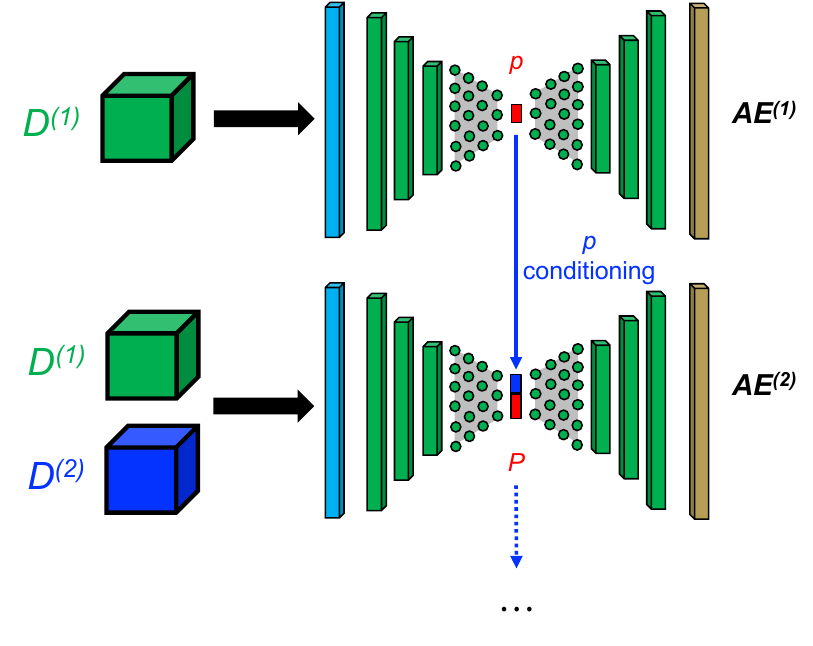}
    \caption{Schematic representation of the training and architecture of the AESTAR for multi-fidelity databases.}
    \label{fig.AESTAR_MF}
\end{figure}

Consider a database consisting of $N$ low-fidelity cases and $M$ high-fidelity cases with $M < N$; we shall refer to these  datasets as $D^{(1)}$ and $D^{(2)}$, respectively. We assume that the high-fidelity cases always correspond to conditions for which also low-fidelity data is available, i.e. $D^{(2)} \in D^{(1)}$.We first train an autoencoder $AE ^{(1)}$ with latent dimension $p$ using $D^{(1)}$; the dimension $p$ is chosen to sufficiently represent the characteristics of the low-fidelity dataset. At this stage, a new training process is initiated for an autoencoder $AE^{(2)}$ with latent space of size $P > p$ and using conditioning to force the first $p$ variables to recover the space corresponding to $AE ^{(1)}$; this  new autoencoder is trained using the multi-fidelity dataset,    $D^{(1)} \bigcup D^{(2)}$. Figure  \ref{fig.AESTAR_MF} illustrates the process.
We shall adopt the AESTAR here to study a database of airfoil flows where the \emph{low}-fidelity data will consist of inviscid  flows, whereas the  \emph{high}-fidelity data  are obtained via RANS viscous simulations.

\section{Case 1:  Inviscid linear theory}
\label{sec.thin}
The incompressible, inviscid  flow  around a two-dimensional airfoil 
immersed in a uniform stream with asymptotic velocity $V_\infty$ is governed by the Laplace equation. In the case of infinitely \emph{thin} airfoils at \emph{small} AoA $\alpha$,
linearized solutions of the Laplace equation are
obtained by superimposing a distribution  of vortices 
along the airfoil chord \cite{Gla47}.
Defining the airfoil geometry as
\begin{equation}
        y=C(x) ,
\end{equation}
    where $C(x)$ is the mean line 
    ($C(0)=C(c)= 0$). 
and defining:
\begin{equation}
    C'(x) =\sum_{n=0}^\infty A_n \cos(n\theta),
\end{equation}
it is
$A_0=\frac{1}{\pi}\int_0^\pi\! \! C'(x) \dif \theta$, and for $n \ge 1:
A_n=\frac{2}{\pi}\int_0^\pi \! \!C'(x) \cos(n\theta) \dif \theta$
with 
\begin{equation}
    \frac{x}{c}=\frac{1}{2}(1-\cos \theta) .
\end{equation}
The kinetic potential is therefore:
\begin{equation}
    \phi=V_\infty (x+\alpha y) +\varphi_\gamma 
\end{equation}
where
\begin{equation}
 \varphi_\gamma= -\frac{1}{2 \pi}\int_0^c\gamma(\xi)\arctan{\frac{y}{x-\xi}} \dif \xi  .
\end{equation}
 $\gamma$ is the circulation per unit length of the vortex 
distribution and is
\begin{equation}
\gamma(\theta)=2 V_\infty \left[ (\alpha -A_0) \cot \frac{\theta}{2}+ \sum_{n=1}^\infty 
A_n \sin(n\theta)\right] \ .
\label{eq.vor}
\end{equation}
The linearized pressure coefficient, through Bernoulli's theorem, is 
\begin{equation}
    C_p(x,0^\pm)=\mp \frac{\gamma(x)}{V_\infty} \ .
\end{equation}
The pressure load along the airfoil is therefore
\begin{equation}
\Delta C_p(x)= C_p(x,0^-) - C_p(x,0^+) =2 \frac{\gamma(x)}{V_\infty} \ ;
\label{eq.Dp}
\end{equation}
its integration along the airfoil provides the celebrated formula of the lift coefficient of an airfoil:
\begin{equation}
    C_l= 2 \pi(\alpha - \alpha_{zl}),
    \label{eg.tacl}
\end{equation}
where $\alpha_{zl}=A_0-A_1/2$ is the zero lift angle.



\begin{figure}
   \begin{subfigure}{1.0\linewidth}
       \includegraphics[width=0.8\textwidth]{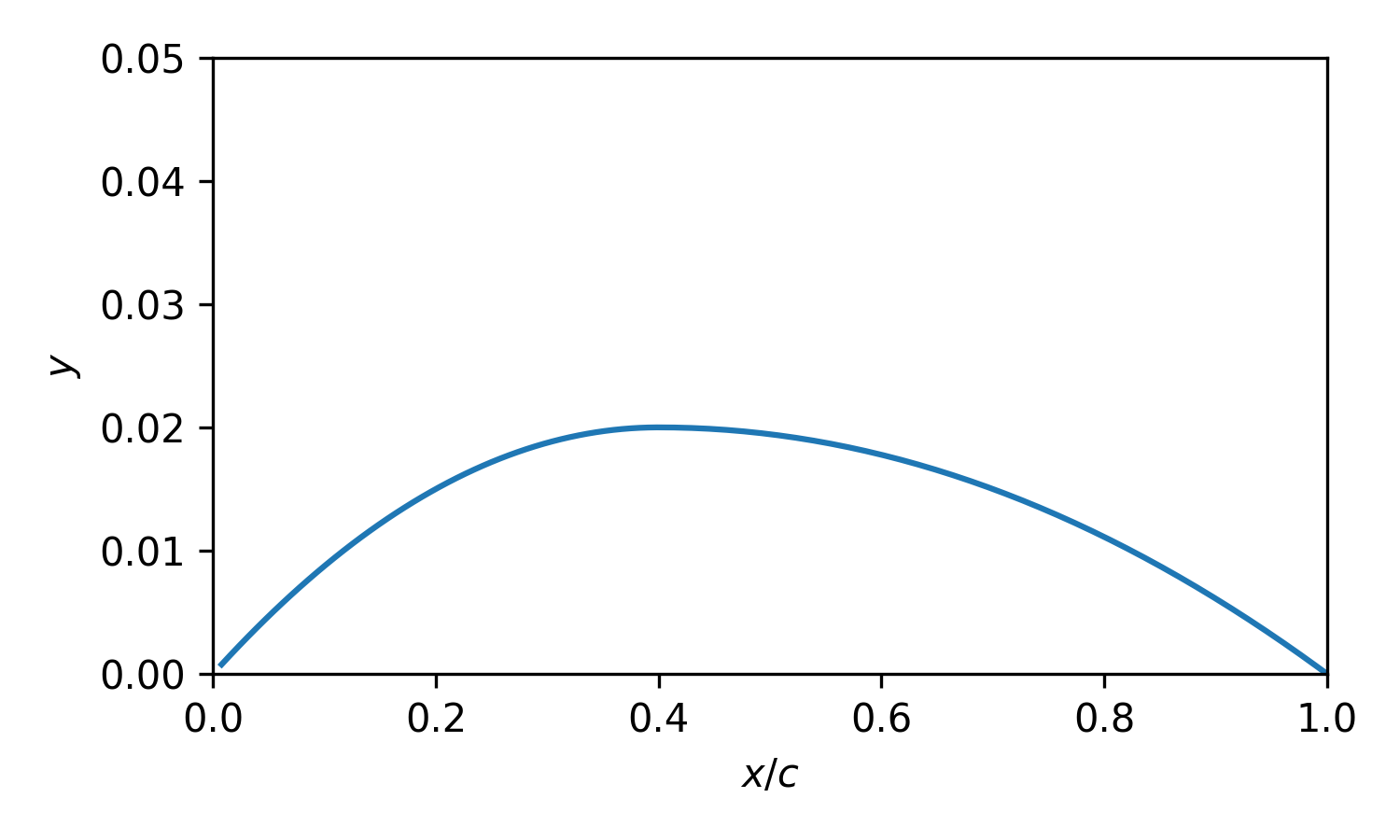}
       \caption{}
   \end{subfigure}
   \hfill
   \begin{subfigure}{1.0\linewidth}
       \includegraphics[width=0.8\textwidth]{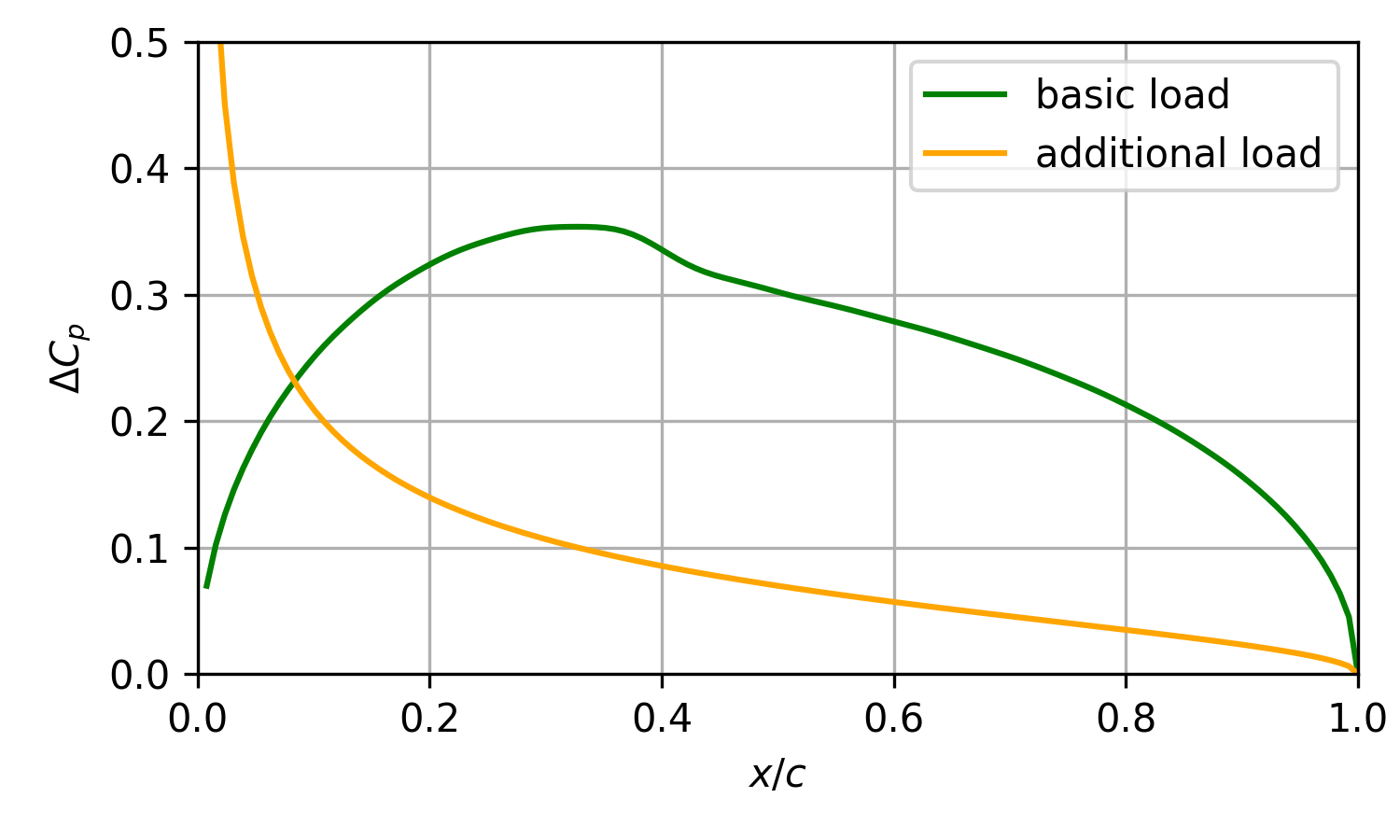}
       \caption{}
   \end{subfigure}
   \caption{(a): NACA 24 mean line. (b): associated basic pressure load   ($A_0 = 0.2574^\circ$) and additional pressure load $(\alpha - A_0 = 1^\circ)$ along the airfoil.}
   \label{fig.basic}
\end{figure}

Eq. \eqref{eq.vor} shows that the load is built-up by two contributions: the additional load, linear function of $(\alpha-A_0)$
($A_0$ is named ideal AoA) and the basic load independent of the AoA and linearly growing with airfoil camber.
Basic and additional load (the latter independent of mean line) are plotted in figure \ref{fig.basic} for the NACA 24 mean line.
Eq. \eqref{eg.tacl} shows that the aerodynamic force varies linearly with the AoA, whereas $\alpha_{zl}$ is a linear function of the airfoil camber.

The database providing the ground truth was built adopting  the NACA X4 mean lines \cite{Abbott1959TheoryOW}. The  geometry of a NACA $D_1D_2$ mean line is given by the equations
\begin{equation}
\begin{split}
    x \le p : & \quad  C(x) = \frac{m}{p^2}\left(2 p x - x^2\right)\\
x > p : & \quad  C(x) = \frac{m}{(1-p)^2}
\left(1-2 p + 2 p x - x^2\right)
\end{split}
\end{equation}
where $m=D_1/100$ is the maximum camber and $p=D_2/10$
is the position of the maximum camber (here $C(x)$ and $x$ are normalized against , the airfoil chord $c$).
The database was built with the NACA 00, 24, 44, 64 and 84 mean lines. For each geometry the pressure load was specified on 257 equally spaced points along the airfoil chord  for the AoAs $\alpha \in [0^\circ, 10^\circ]$
step $1^\circ$, for a total of 55 data points. 

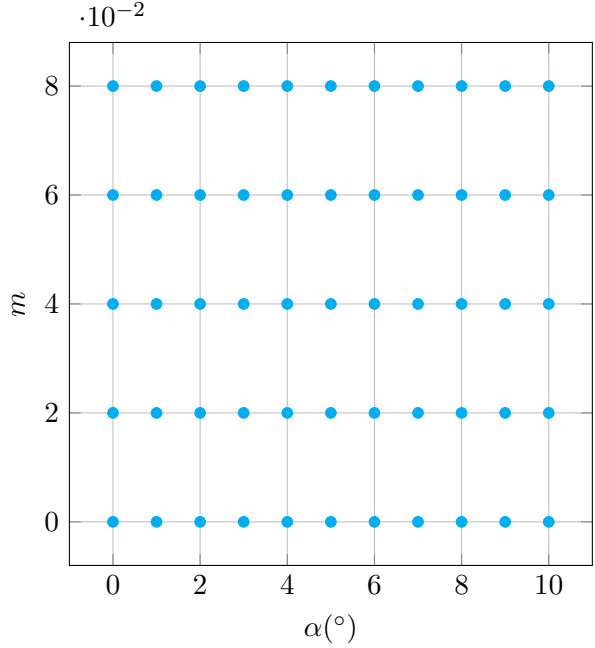
\begin{figure}
\centering
\begin{tikzpicture}
\begin{axis}[width=8.5cm,height=8.5cm,
grid,xlabel=$\alpha(^\circ)$, ylabel=$m$]
\addplot[only marks,cyan,mark=*]
 table [x expr=\thisrowno{0}, y expr=\thisrowno{1}] {plots/PLS_ThinAF.txt}; 
\end{axis}
\end{tikzpicture}
       \caption{PLS of the thin airfoil database.}
   \label{fig.physls1}
\end{figure}

The objective of the latent representation is to retain the information required by the decoder to reconstruct the input.
A   PLS is a low-dimensional parameterization of the database constructed from known physical parameters or observables that uniquely characterize the physical states considered.
The pressure load distribution  is identified by $\alpha$ and maximum camber $m$ all reported in the diagram shown in figure \ref{fig.physls1} for the present database. It is a regular, linear and orthogonal two-dimensional representation of all data points; it is therefore a PLS representing our database with latent space dimension $P=2$.
Each latent space point $(\alpha, m)$  provides the pressure load for each data point through eq. \eqref{eq.Dp}:  $\vect{\Delta p}=\vect{\Delta p}(\alpha,m)$  is therefore an analytical (and not data-driven) decoder. 

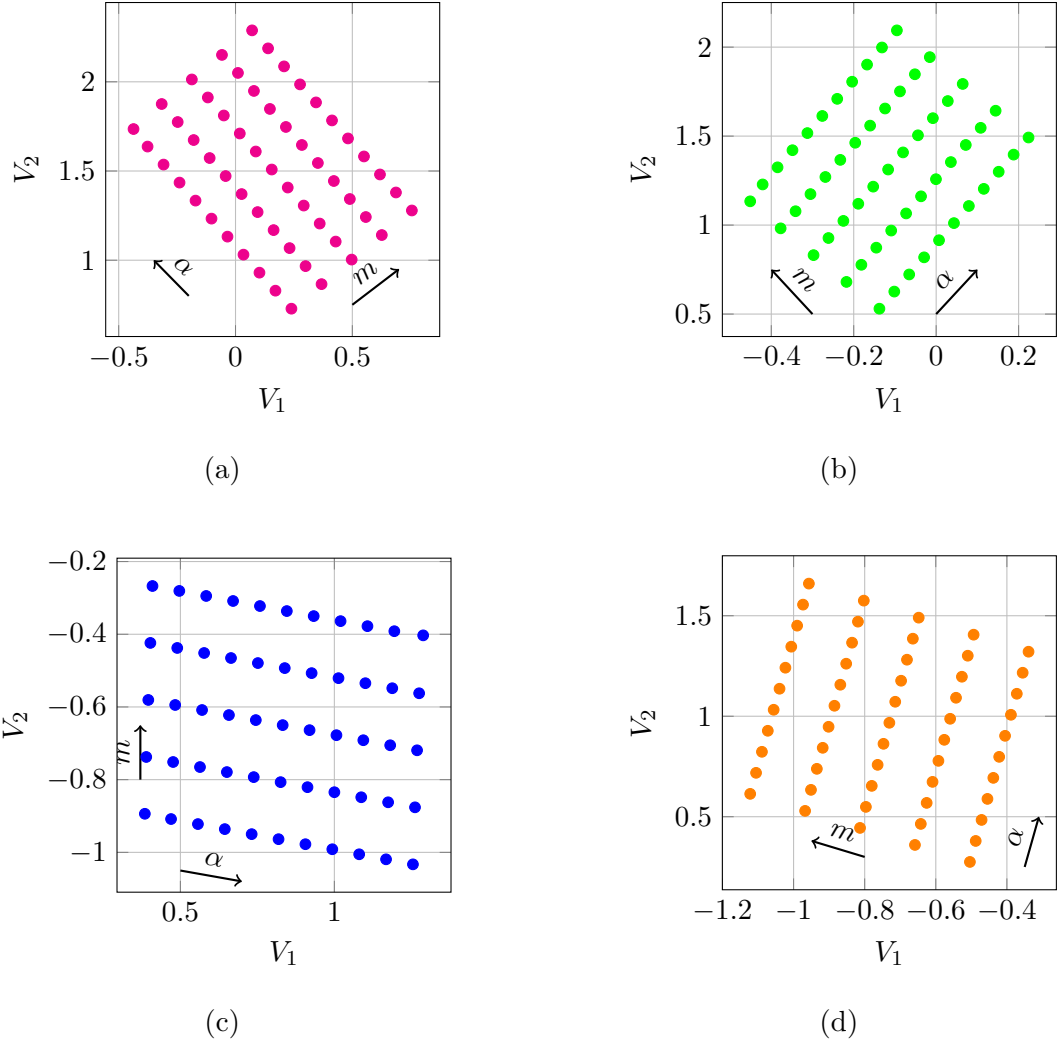
\begin{figure*}
\centering
\begin{subfigure}{0.45\textwidth}
\begin{tikzpicture}
\begin{axis}[width=6cm,height=6.0cm,
grid,xlabel=$V_1$, ylabel=$V_2$]
\addplot[only marks,magenta,mark=*]
 table [x expr=\thisrowno{1}, y expr=\thisrowno{2}] {plots/ThinAF_V1_V2_a.txt}; 
 \draw[->, thick]
    (axis cs:-0.2,0.8) --
    (axis cs:-0.35,1.0)
    node[midway, above, sloped] {$\alpha$};
 \draw[->, thick]
    (axis cs:0.5,0.75) --
    (axis cs:0.7,0.95)
    node[midway, above, sloped] {$m$};
\end{axis}
\end{tikzpicture}
\caption{}
\end{subfigure}
\hspace{5mm}
\begin{subfigure}{0.45\textwidth}
\begin{tikzpicture}
\begin{axis}[width=6cm,height=6.0cm,
grid,xlabel=$V_1$, ylabel=$V_2$]
\addplot[only marks,green,mark=*]
 table [x expr=\thisrowno{1}, y expr=\thisrowno{2}] {plots/ThinAF_V1_V2_b.txt}; 
 \draw[->, thick]
    (axis cs:0.0,0.5) --
    (axis cs:0.1,0.75)
    node[midway, above, sloped] {$\alpha$};
 \draw[->, thick]
    (axis cs:-0.3,0.5) --
    (axis cs:-0.4,0.75)
    node[midway, above, sloped] {$m$};
\end{axis}
\end{tikzpicture}
\caption{}
\end{subfigure}

\vspace{5mm}
\begin{subfigure}{0.45\textwidth}
\begin{tikzpicture}
\begin{axis}[width=6cm,height=6.0cm,
grid,xlabel=$V_1$, ylabel=$V_2$]
\addplot[only marks,blue,mark=*]
 table [x expr=\thisrowno{1}, y expr=\thisrowno{2}] {plots/ThinAF_V1_V2_c.txt}; 
 \draw[->, thick]
    (axis cs:0.5,-1.05) --
    (axis cs:0.7,-1.08)
    node[midway, above, sloped] {$\alpha$};
 \draw[->, thick]
    (axis cs:0.37,-0.8) --
    (axis cs:0.37,-0.65)
    node[midway, above, sloped] {$m$};
\end{axis}
\end{tikzpicture}
\caption{}
\end{subfigure}
\hspace{5mm}
\begin{subfigure}{0.45\textwidth}
\begin{tikzpicture}
\begin{axis}[width=6cm,height=6.0cm,
grid,xlabel=$V_1$, ylabel=$V_2$]
\addplot[only marks,orange,mark=*]
 table [x expr=\thisrowno{1}, y expr=\thisrowno{2}] {plots/ThinAF_V1_V2_d.txt};
 \draw[->, thick]
    (axis cs:-0.35,0.25) --
    (axis cs:-0.31,0.50)
    node[midway, above, sloped] {$\alpha$};
 \draw[->, thick]
    (axis cs:-0.8,0.3) --
    (axis cs:-0.95,0.38)
    node[midway, above, sloped] {$m$};
\end{axis}
\end{tikzpicture}
\caption{}
\end{subfigure}
\caption{Thin airfoil database analysis. Vanilla AE, latent space variables $V_1$, $V_2$ as obtained by 4 different trainings.}
     \label{fig.TAtrainingV1V2}
    \label{fig.TAFaetrains}
\end{figure*}

In a linear problem, an unconstrained  autoencoder is not guaranteed to learn a linear latent representation. Nonlinear encoder–decoder pairs can represent the same linear system through nonlinear latent coordinates. It is therefore interesting to look here at the answer given by the application of a vanilla AE with  latent space dimension $P=2$. 
Figure \ref{fig.TAtrainingV1V2} reports the DDLS $(V_1,V_2)$ as obtained by 4 different random trainings (further trainings, when converged, always showed the same structure). Each training, starting with a random choice of the data points and NN weights,  provided a different representation in the latent space, however the structures are all similar and what is more interesting, also similar to the diagram of the PLS already shown in figure \ref{fig.physls1}.
There is a clear ordering with the physical parameters even if directions and orientations, such as the adopted metric, can be different in each training. 
The linearity of the ground truth respect to both AoA and airfoil camber is  reproduced.
As in the case of the diagram of figure \ref{fig.physls1}, each point $(V_1, V_2)$ provides, through the AE decoder, the pressure load of the associated data point.

    
Giral et al.\cite{aerojepa2026} suggested to quantify the connection of the DDLS with physical parameters by ridge regression (see appendix \ref{app.ridge} for a brief description); they assume the correlation  sufficient  if the coefficient of determination $R^2 > 0.85$.
Table \ref{tab.ThinAF} shows $R^2$   between the 4 DDLSs and the parameters $\alpha$ and $m$. The regularization parameter was selected by cross-validation. 
$R^2 = 1$ for all training in practice, it enforces the correlation between the PLS and all obtained DDLSs.
From a mathematical viewpoint, this result ensures that 
a PLS data point can be obtained by a linear combination of DDLS points.

Incidentally, since in the frame of the linearized inviscid theory lift is a linear function of the AoA, see eq. \eqref{eg.tacl}, in all DDLS proposed $\alpha$ can be replaced with lift. In this case the DDLS shows that lift also linearly vary with the camber $m$ of the airfoil which is one of the fundamental results of the airfoil theory developed one century ago and correctly represented in the DDLS.

All obtained latent spaces look  a rotation, reflection or deformation of the 
PLS, that is therefore just  
\emph{one of the  possible latent space representations of  the database}. 
Although the coordinate geometry varies among trainings, the learned representations preserve the organization of the underlying physical manifold and the linearity of the physics is reproduced by their shape: a strict trapezoid.
Procrustes analysis (PA) \cite{Schonemann1966} or general affine transformation (GAT) \cite{Hartley2004} can be used to quantify the geometric relations between PLS and DDLS. 
 The Orthogonal PA  determines the
orthogonal transformation that minimizes the discrepancy between two
configurations, whereas a less restrictive comparison can be obtained using the GAT, which may additionally
include anisotropic scaling and shear. A short summary of both method is reported in  appendix
\ref{app.procrustes_affine}. 
The normalized errors  of PA $(E_{P})$
and GAT $(E_{A})$ are also
presented in table \ref{tab.ThinAF} for the shown trainings. 
The $R^2$ value proposed in the table says whether a physical quantity is linearly decodable from the latent space.
$E_P$ versus $E_A$ says what kind of linear map it is: a pure rotation or reflection, or one that also involves anisotropic scaling and shear
The lower the error, the stronger is the geometric relation between PLS and DDLS.
While the normalized error of PA is rather high ($E_{P} \approx 20\%$), as could be expected since PA does not allow for shear deformation, the error of the GAT is extremely low ($E_{A} < 0.3\%$). 
DDLS geometry is not simply related to the PLS by translation, rotation/reflection, and uniform scaling, but it is much closer to the PLS after allowing the additional degrees of freedom of a general affine transformation, particularly anisotropic scaling and shear: the DDLSs can be therefore regarded as affinely equivalent to the PLS: \emph{even if  converged AE trainings
admit infinite different solutions, and the PLS can be intepreted as one of them, at least for this test, from a geometrical view point they are all affinely equivalent.}

\begin{figure}
    \centering
    \begin{tikzpicture}
    \begin{axis}[width=8.5cm,height=8.5cm,
    grid,xlabel=PC$_1$, ylabel=PC$_2$]
    \addplot[only marks,cyan,mark=*]
     table [x expr=\thisrowno{1}, y expr=\thisrowno{2}] {plots/ThinAF_V1_V2_PCA.txt};
    \draw[->, thick]
    (axis cs:-1.25,-0.5) --
    (axis cs:-0.75,-0.65)
    node[midway, above, sloped] {$\alpha$};
 \draw[->, thick]
    (axis cs:-1.25,0.25) --
    (axis cs:-1,0.5)
    node[midway, above, sloped] {$m$};
    \end{axis}
    \end{tikzpicture}
    \caption{Thin airfoil database analysis. PCA, principal components PC$_1$, PC$_2$.}
    \label{fig:PCA_ThinAF}
\end{figure}
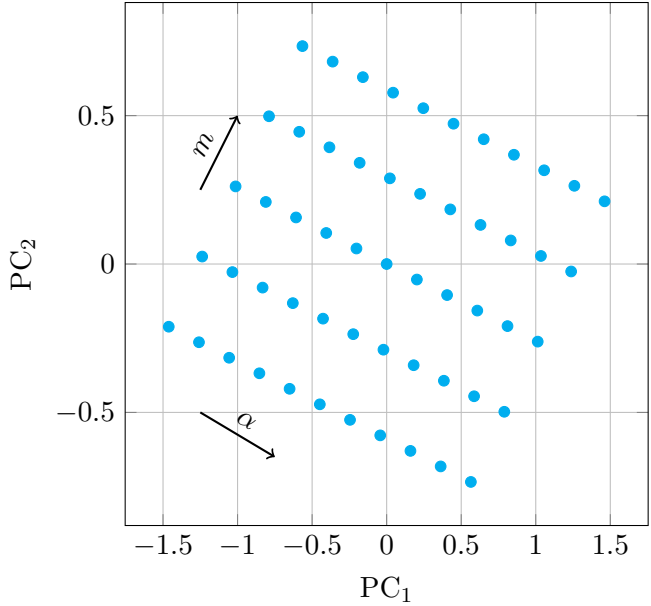

The close relationship between AEs and PCA is well established \cite{Kramer1991}. Accordingly, we also carried out PCA on the same dataset, and figure \ref{fig:PCA_ThinAF} presents the results in the principal component plane. The PCA result can be interpreted as a rotated and rescaled version of the PLS, which supports the equivalence between AE and PCA for this linear case.

\begin{table}[]
    \centering
    \begin{tabular}{|c|c|c|c|c|}
    \hline
    \hline
    DDLS & $\alpha$ & $m$ & $E_{P}$ & $E_{A}$\\
  \hline
  \hline
  Training (a) & 0.9999 & 0.9999 &0.1632 & 0.0027\\
 Training  (b) & 0.9999 & 0.9999 &0.3526& 0.0032\\
 Training  (c) & 0.9999 & 0.9999 &0.2029& 0.0029\\
 Training  (d) & 1.0000 & 1.0000 &0.2241 & 0.00009 \\
  \hline
  \hline
    \end{tabular}
    \caption{Thin airfoil database analysis. 
    Results of 4 trainings.
    $R^2$ values of ridge regression
    on $\alpha$ and $m$ 
    and normalized errors of PA and GAT of DDLS with PLS ($\alpha, m$).}
    \label{tab.ThinAF}
\end{table}

\section{Case 2: Joukowski airfoil}

Exact (non linearized) analytical solutions of the inviscid incompressible flow around airfoils can be obtained through conformal mapping; the most famous solution is the Joukowski airfoil transformation \cite{Mil1966}.
The complex potential $W=\varphi + i \psi$
of the flow around a circular cylinder of radius $R$ placed in the origin of a complex plane $z''=x''+i y''$ and characterized by a circulation $\Gamma$ (positive  clock wise)
is:
\begin{equation}
\label{eq:Wz}
    W=V_\infty \left(z''+\frac{R^2}{z''}\right)+ i \frac{\Gamma}{2 \pi} \ln\frac{z''}{R} \ .
\end{equation}
The flow at an AoA $\alpha$ around the same cylinder is obtained 
by the conformal mapping $z'=z''\mathrm{e}^{i\alpha}$. The   case cylinder centered in $z_0=1-R \mathrm{e}^{-i\beta}$ with $R>1$ requires the additional  mapping $z=z'+z_0$.

The Joukowski conformal mapping
\begin{equation}
\label{eq:Jt}
    \zeta= z + \frac{1}{z}
\end{equation}
transforms the circle in the $z$ plane in an airfoil in the plane $\zeta=\xi+i \eta$. 
The airfoil shape is identified by the two parameters
$R$ and $\beta$. $R$ controls the airfoil thickness while  $\beta$ grows with the airfoil camber. 

The velocity components $u,v$ can be  computed by taking the derivative of complex potential:
\begin{equation}
    \frac{dW}{d\zeta}=u -i v
\end{equation}
and finally the pressure coefficient by Bernoulli's theorem.
Among the infinite solutions obtainable by varying
the circulation $\Gamma$, the only acceptable from a physical viewpoint is selected imposing the continuity of the velocity field at the airfoil trailing edge (Kutta condition):
\begin{equation}
    \Gamma = 4 \pi V_\infty R \sin(\alpha+\beta) \, .
\end{equation}
Therefore, the Kutta-Joukowski theorem allows for the calculation of  the lift coefficient:
\begin{equation}
    C_l=8 \pi \frac{R}{c} \sin (\alpha+ \beta) \ ,
    \label{eq.clabeta}
\end{equation}
where  $c$ is the airfoil chord.
For symmetrical airfoils ($\beta=0$), the lift is in particular proportional to the sine of the angle of attack. 

The first answer we are looking for in this section is to verify if a DDLS can provide a \emph{quantitative} physical information of the flow around the airfoil.
As in the previous case, we first look for a PLS with an analytical  \emph{decoder} given by the  solution of the Joukowski airfoil.

\begin{figure*}
\centering
\begin{subfigure}{0.45\textwidth}
\centering
\begin{tikzpicture}
\pgfplotsset{every axis legend/.append style={at={(1,1)},anchor=north east}}
\begin{axis}[width=5.5cm,height=7.5cm,
grid,xlabel=$x$, ylabel=$C_p$, xmin=-2.1, xmax=2.1,ymin=-13.0,ymax=1.0, y dir=reverse]
\addplot[thick,red,solid,smooth]
 table [x expr=\thisrowno{0}, y expr=\thisrowno{2}] {plots/Jou1DCpa1.0R1.10b0.0AoA0.0.txt}; 
\addplot[thick,black,solid,smooth]
 table [x expr=\thisrowno{0}, y expr=\thisrowno{2}] {plots/Jou1DCpa1.0R1.10b0.0AoA5.0.txt}; 
 \addplot[thick,blue,solid,smooth]
 table [x expr=\thisrowno{0}, y expr=\thisrowno{2}] {plots/Jou1DCpa1.0R1.10b0.0AoA10.0.txt}; 
 \addplot[thick,green,solid,smooth]
 table [x expr=\thisrowno{0}, y expr=\thisrowno{2}] {plots/Jou1DCpa1.0R1.10b0.0AoA15.0.txt}; 
 \legend{\tiny $\alpha=0^\circ$, \tiny $\alpha=5^\circ$, \tiny $\alpha=10^\circ$,\tiny $\alpha=15^\circ$}
\end{axis}
\end{tikzpicture}
\caption{}
\end{subfigure}
\hspace{3mm}
\begin{subfigure}{0.45\textwidth}
\centering
\begin{tikzpicture}
\begin{axis}[width=5.5cm,height=7.5cm,
grid,xlabel=$\alpha \ (^\circ)$, ylabel=$C_{p_{min}}$, xmin=0, xmax=90, ymax=0]
\addplot[ultra thick,cyan,solid,smooth]
 table [x expr=\thisrowno{0}, y expr=\thisrowno{1}] {plots/Jou_alpha_Cpmin.txt}; 
\end{axis}
\end{tikzpicture}
\caption{}
\end{subfigure}
\caption{Joukowski airfoil $R=1.1$, $\beta=0$. (a) surface pressure coefficients at different angles of attack. (b) Minimum pressure coefficient vs AoA. }
    \label{fig.JouCp1D}
\end{figure*}
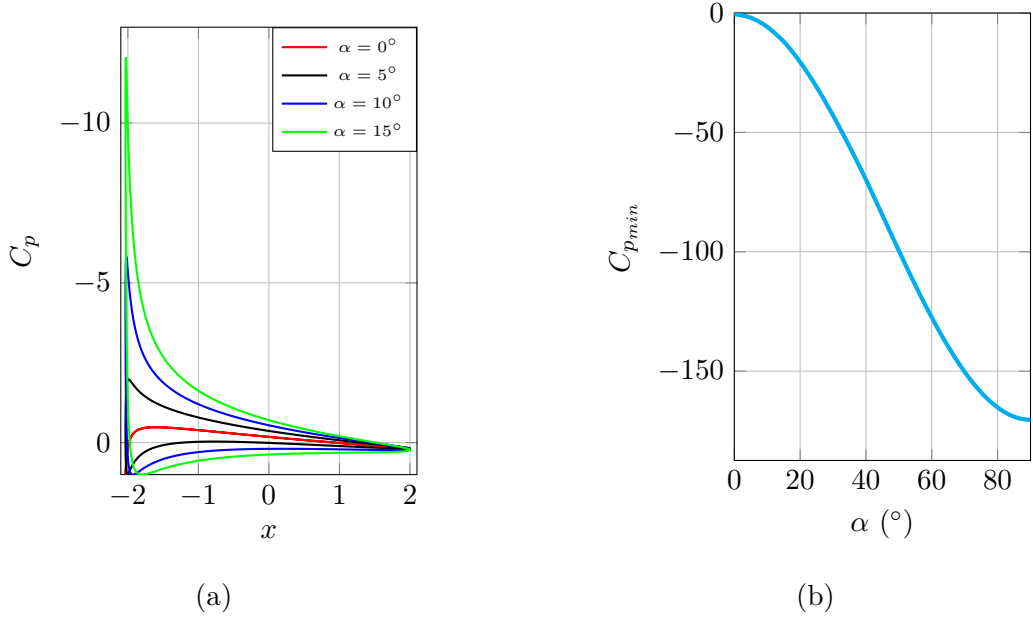

Figure \ref{fig.JouCp1D}(a) shows the surface pressure coefficients for a symmetric Joukowski airfoil at different angles of attack. 
The minimum pressure peak on the airfoil (and in the complete flow) is a monotonic function of the AoA and is displayed against the AoA in figure \ref{fig.JouCp1D}(b). For a given airfoil, known the pressure field,
the minimum pressure value \emph{uniquely} identifies the associated flow.

\begin{figure*}
\centering
\begin{subfigure}{0.45\textwidth}
\centering
\includegraphics[width=0.8\textwidth]{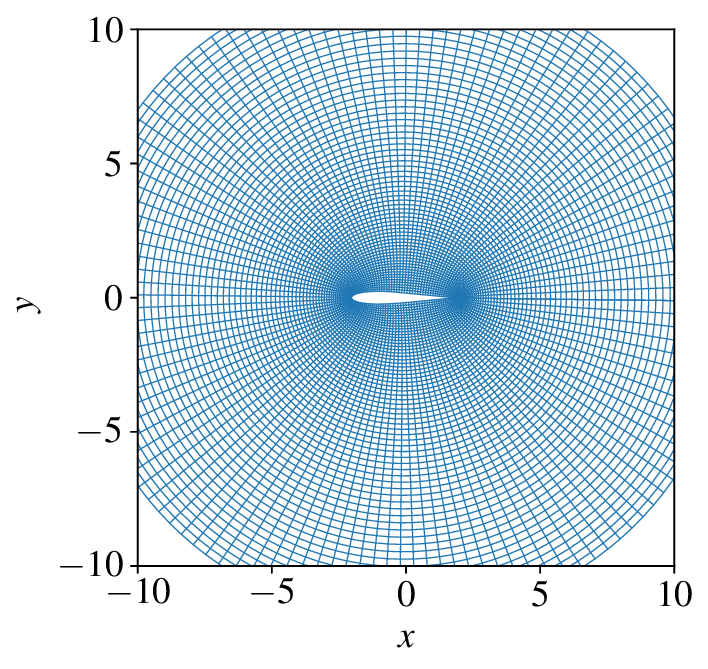}
\caption{$129 \times 65$ grid where data are stored.
    }
\end{subfigure}
\hspace{3mm}
\begin{subfigure}{0.45\textwidth}
\centering
\includegraphics[width=0.95\textwidth]{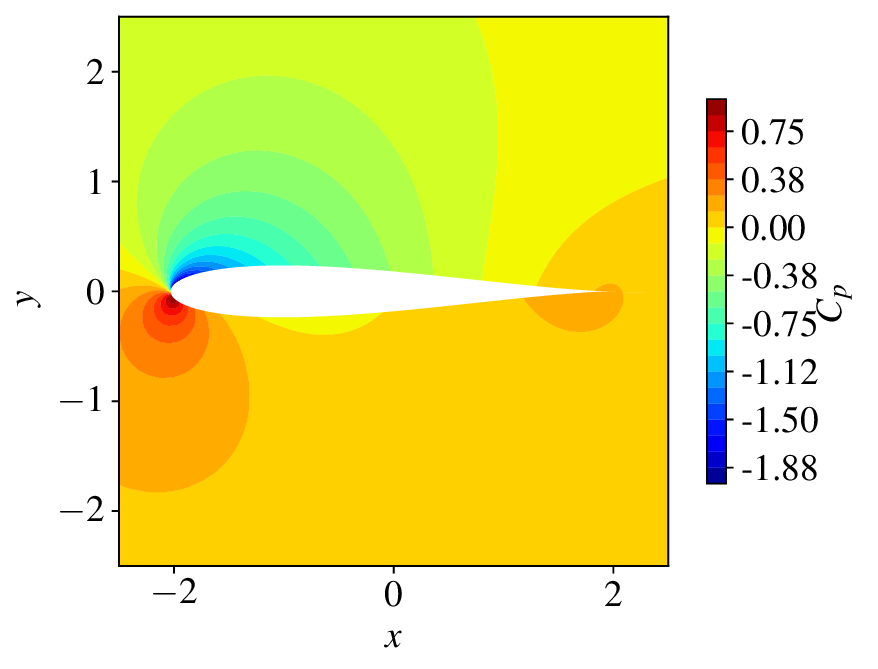}
\caption{Pressure coefficient at $\alpha=5^\circ$.}
\end{subfigure}
    \caption{Joukowski airfoil $R=1.1$ and $\beta=0$.
    }
    \label{fig.Joukowski_grid}
\end{figure*}

Even if the Joukowski flows are physically relevant  only for small $\alpha$,
we built a data base containing the pressure field of the  airfoil  $R=1.1$, $\beta=0$, for $\alpha \in[0^\circ, 90^\circ]$, step $1^\circ$, in order to increase the complexity of this \emph{toy} problem.
The pressure coefficient was sampled in 129 grid points around the airfoil and 65 points in normal direction as shown in figure \ref{fig.Joukowski_grid}. 
Each image of the database was built with $129 \times 65$ pixels and contained just one channel: the pressure coefficient.
The minimum pressure value as plotted in figure 
\ref{fig.JouCp1D}(b) is therefore the wanted PLS representation of the database with latent space dimension $P=1$.

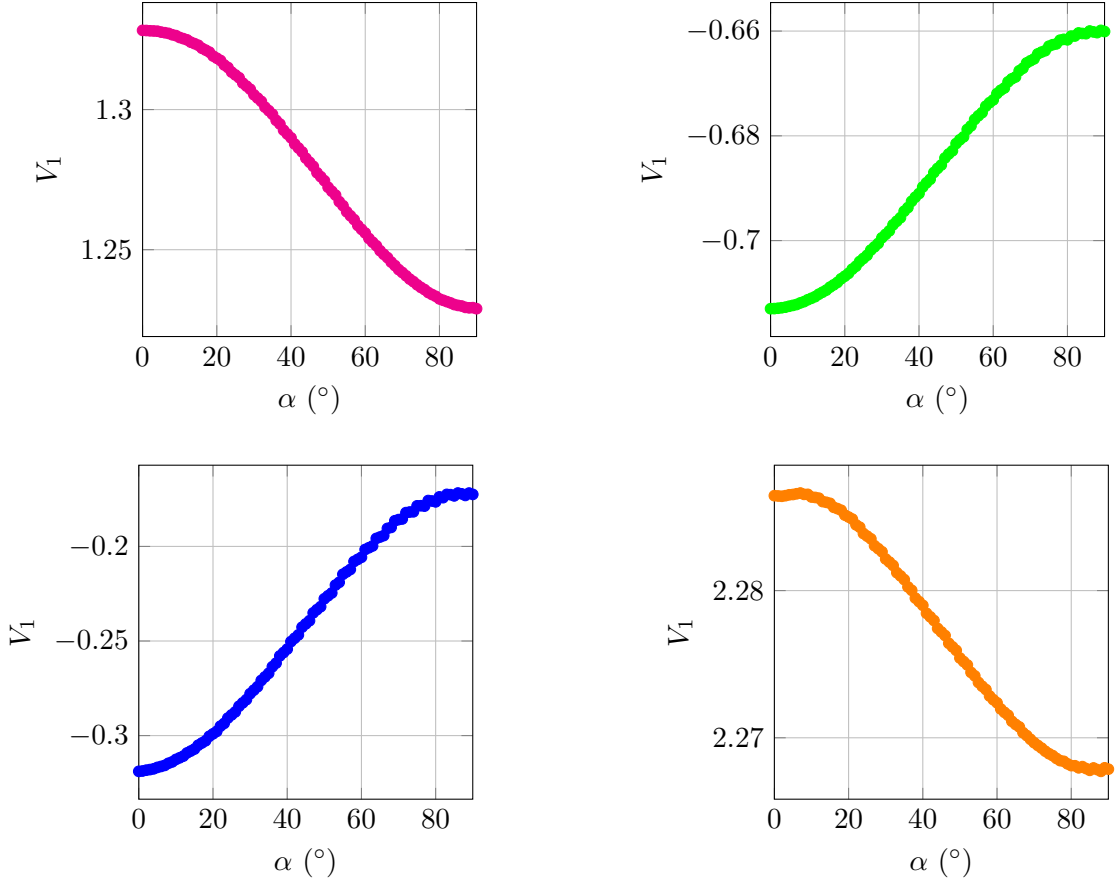
\begin{figure*}
\centering
\begin{subfigure}{0.45\textwidth}
\begin{tikzpicture}
\begin{axis}[width=6cm,height=6.0cm,
grid,xlabel=$\alpha \ (^\circ)$, ylabel=$V_1$, xmin=0, xmax=90]
\addplot[only marks,magenta,mark=*]
 table [x expr=\thisrowno{0}, y expr=\thisrowno{1}] {plots/Jou_alpha_V1.txt}; 
\end{axis}
\end{tikzpicture}
\end{subfigure}
\hspace{5mm}
\begin{subfigure}{0.45\textwidth}
\begin{tikzpicture}
\begin{axis}[width=6cm,height=6.0cm,
grid,xlabel=$\alpha \ (^\circ)$, ylabel=$V_1$, xmin=0, xmax=90]
\addplot[only marks,green,mark=*]
 table [x expr=\thisrowno{0}, y expr=\thisrowno{1}] {plots/Jou_alpha_V1_b.txt}; 
\end{axis}
\end{tikzpicture}
\end{subfigure}

\vspace{5mm}
\begin{subfigure}{0.45\textwidth}
\begin{tikzpicture}
\begin{axis}[width=6cm,height=6.0cm,
grid,xlabel=$\alpha \ (^\circ)$, ylabel=$V_1$, xmin=0, xmax=90]
\addplot[only marks,blue,mark=*]
 table [x expr=\thisrowno{0}, y expr=\thisrowno{1}] {plots/Jou_alpha_V1_c.txt}; 
\end{axis}
\end{tikzpicture}
\end{subfigure}
\hspace{9mm}
\begin{subfigure}{0.45\textwidth}
\begin{tikzpicture}
\begin{axis}[width=6cm,height=6.0cm,
grid,xlabel=$\alpha \ (^\circ)$, ylabel=$V_1$, xmin=0, xmax=90]
\addplot[only marks,orange,mark=*]
 table [x expr=\thisrowno{0}, y expr=\thisrowno{1}] {plots/Jou_alpha_V1_d.txt}; 
\end{axis}
\end{tikzpicture}
\end{subfigure}
\caption{Joukowski airfoil $R=1.1$, $\beta=0$.
Latent variable $V_1$ against AoA as obtained by four different trainings.}
    \label{fig.JouV1s}
\end{figure*}

The latent variables $V_1$ as obtained during 
4  trainings of the vanilla AE with $P=1$,
again with random data choice and weight settings,
are plotted  in figure \ref{fig.JouV1s} against the AoA. The shape of the curves are impressively similar to the plot of the minimum pressure. It is an effective example of physical imprinting.

\begin{table}
    \centering
    \begin{tabular}{|l|c|}
    \hline
    \hline
       \textbf{Parameter}  & \textbf{Range} \\
         \hline
         \hline
       Number of convolutions   & 4-8 \\
        Number of fully connected & 1-4\\
       Kernel size  & 2-5 \\
       Kernel stride  & 1-3 \\
       Max pooling  & On-Off \\
       Max pooling size   & 2-4 \\
       Max pooling stride   & 2-3 \\
       Drop out probability  & 0\%-25\% \\
         \hline
         \hline
    \end{tabular}
    \caption{Variability of the architecture ensemble in training of Joukowski database 1.}
    \label{tab.uq}
\end{table}

\begin{figure}
\centering
\includegraphics[scale=0.45]{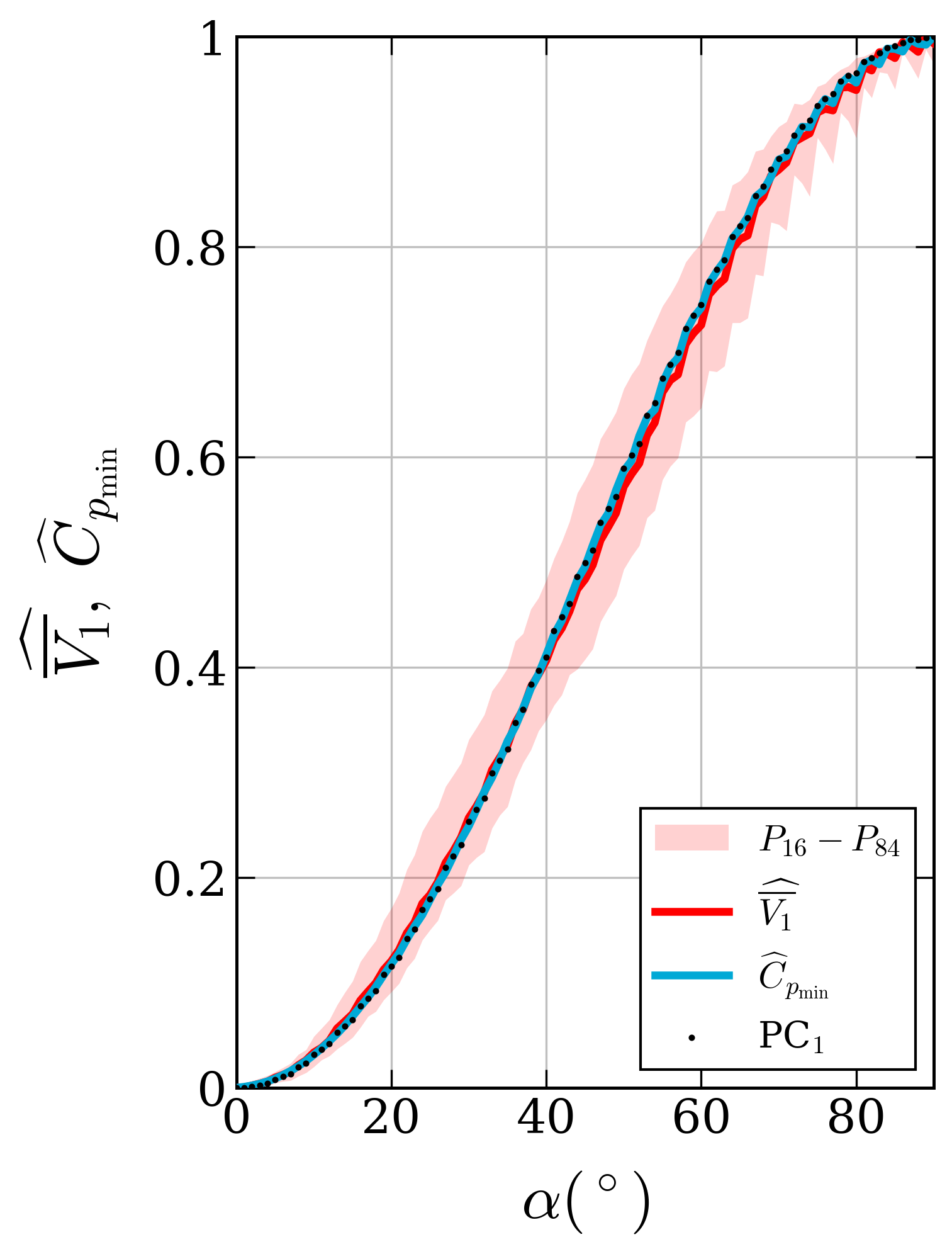}
\caption{Joukowski airfoil $R=1.1$, $\beta=0$.
Normalized $C_{p_{\rm min}}$compared with the median normalized latent variable
($\hat{\bar{V_1}}$) and PCA ($\mathrm{PC_1}$).
}
    \label{fig.JouV1}
\end{figure}

The comparison between the normalized latent variables $\hat{V}_1$ obtained from the different trainings and the normalized minimum pressure coefficient $\hat{C}{p_{\min}}$ is shown in Figure \ref{fig.JouV1}. The training procedure is performed using an ensemble of 100 AEs with varying architectures \cite{saettaUQ}, and the variability across the architectural ensemble is reported in Table \ref{tab.uq}. As shown in the figure, the median latent variable obtained from the ensemble closely matches the PLS prediction while the shaded band represents the 16th–84th percentile interval of the latent-variable ensemble. The figure also shows the normalized PCA result; it is in agreement with both $C_{p_{min}}$ and AE output. 
The first principal component shows the same dependence on $\alpha$. The representation therefore reflects how variance is distributed in the data: the leading-edge suction peak dominates the variability of the pressure field, and both linear and nonlinear compression retain it as the primary coordinate. In this sense the physics is imprinted in the data and preserved by the compression.
This result provides a neat answer to the first question posed in this section: a DDLS can effectively provide a \emph{quantitative} representation of a physical response, in this case the minimum pressure value in the flow.

Quite interestingly, the encoder used the available channel data type (pressure), even if there are  many scalar functions of the flow field that vary monotonically with $\alpha$ ($\alpha$ itself for instance) and could parameterize this one-dimensional manifold.
 On the contrary to what observed by Kang et al. for instance, in this case the DDLSs are correlated with a physical \emph{response} even quantitatively, an output of the system
 and not to a physical \emph{parameter} defining the system, as could be the AoA.

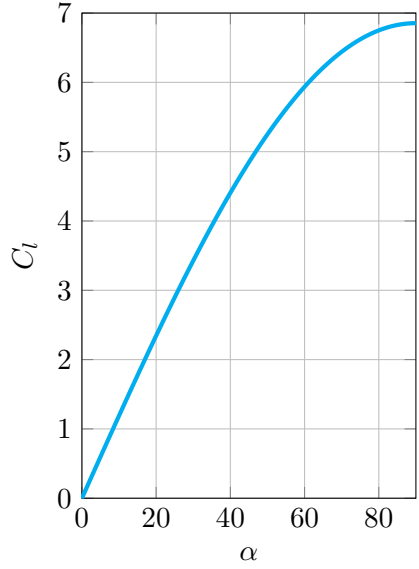
\begin{figure}
    \centering
    \begin{tikzpicture}
    \begin{axis}[height=8cm, width=6cm, grid, xmin=0, xmax = 90,ymin=0,ymax=7,xlabel=$\alpha$, ylabel=$C_l$]
    \addplot[ultra thick,solid, color=cyan, mark size = 3pt]
    table [x expr=\thisrowno{0}, y expr=\thisrowno{1}] {plots/ClaR11.txt};
    \end{axis}
    \end{tikzpicture}
    \caption{Joukowski airfoil $R=1.1$, $\beta=0$. Lift coefficient against AoA.}
    \label{fig.clabeta0}
\end{figure}

A new database was created with two additional channels by adding the cartesian velocity components $u,v$  to the pressure coefficient.
We expect that  the adoption of non homogeneous data (pressure and velocities) could imply a different DDLS representation.
In particular,  adopting a latent space dimension $P=2$ we can uniquely identify a database point  by  AoA and  lift coefficient through eq. \eqref{eq.clabeta} that for symmetric airfoil reduces to
\begin{equation}
        C_l=8 \pi \frac{R}{c} \sin \alpha \ .
        \label{eq.clabeta0}
\end{equation}
This equation, plotted in figure  \ref{fig.clabeta0} in the plane $(\alpha,C_l)$
is an alternative PLS representation of the database with latent space dimension $P=2$. 

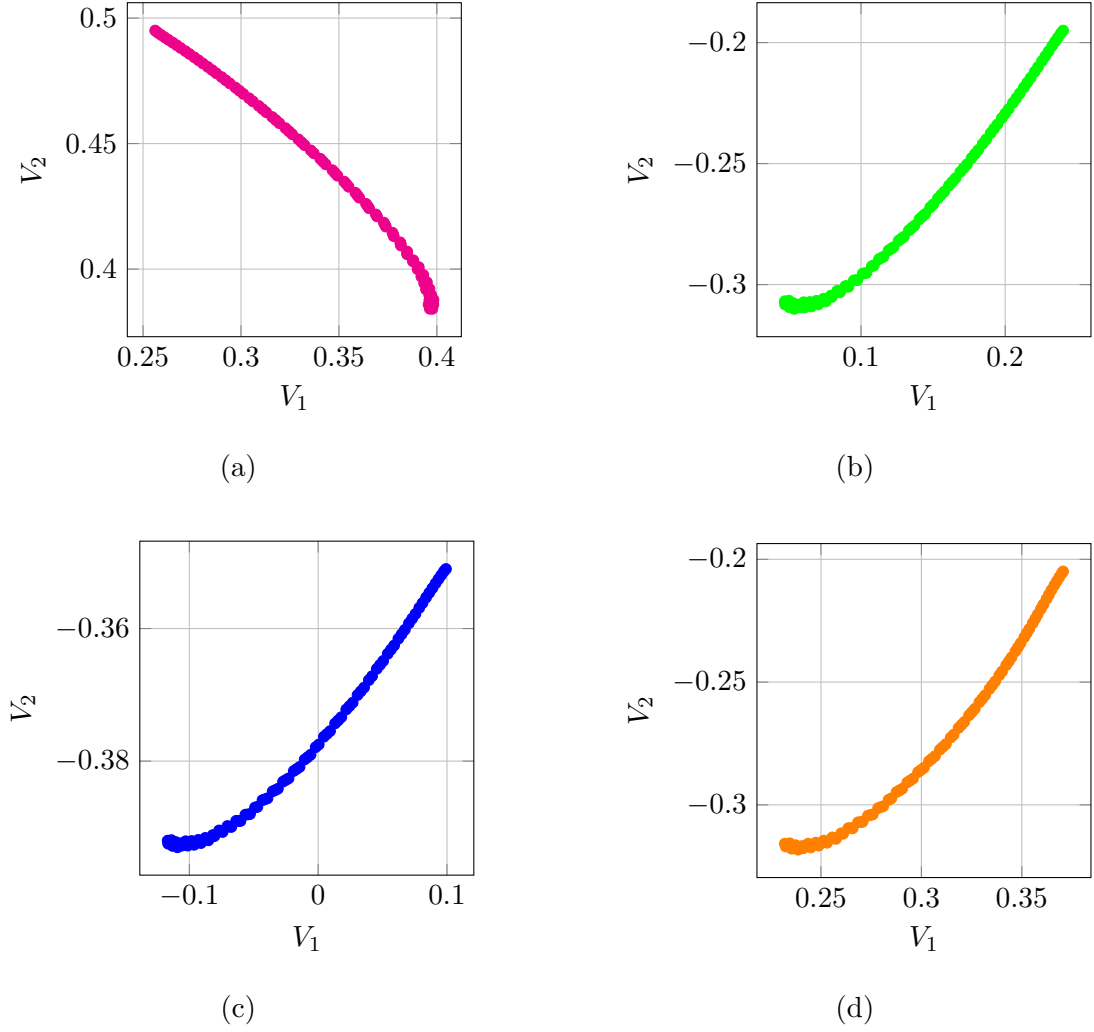
\begin{figure*}
\centering
\begin{subfigure}{0.45\textwidth}
\begin{tikzpicture}
\begin{axis}[width=6cm,height=6.0cm,
grid,xlabel=$V_1$, ylabel=$V_2$]
\addplot[only marks,magenta,mark=*]
 table [x expr=\thisrowno{1}, y expr=\thisrowno{2}] {plots/Jou_alpha_V1_V2_a.txt}; 
\end{axis}
\end{tikzpicture}
\caption{}
\end{subfigure}
\hspace{5mm}
\begin{subfigure}{0.45\textwidth}
\begin{tikzpicture}
\begin{axis}[width=6cm,height=6.0cm,
grid,xlabel=$V_1$, ylabel=$V_2$]
\addplot[only marks,green,mark=*]
 table [x expr=\thisrowno{1}, y expr=\thisrowno{2}] {plots/Jou_alpha_V1_V2_b.txt}; 
\end{axis}
\end{tikzpicture}
\caption{}
\end{subfigure}

\vspace{5mm}
\begin{subfigure}{0.45\textwidth}
\begin{tikzpicture}
\begin{axis}[width=6cm,height=6.0cm,
grid,xlabel=$V_1$, ylabel=$V_2$]
\addplot[only marks,blue,mark=*]
 table [x expr=\thisrowno{1}, y expr=\thisrowno{2}] {plots/Jou_alpha_V1_V2_e.txt}; 
\end{axis}
\end{tikzpicture}
\caption{}
\end{subfigure}
\hspace{5mm}
\begin{subfigure}{0.45\textwidth}
\begin{tikzpicture}
\begin{axis}[width=6cm,height=6.0cm,
grid,xlabel=$V_1$, ylabel=$V_2$]
\addplot[only marks,orange,mark=*]
 table [x expr=\thisrowno{1}, y expr=\thisrowno{2}] {plots/Jou_alpha_V1_V2_d.txt}; 
\end{axis}
\end{tikzpicture}
\caption{}
\end{subfigure}
\caption{Joukowski airfoil $R=1.1$, $\beta=0$.
Latent space variables $V_1$, $V_2$ as obtained by four different trainings.}
    \label{fig.jouV2V1}
\end{figure*}

The two-dimensional DDLS as obtained by 4 random trainings of the vanilla AE with $P=2$, are presented in figure
\ref{fig.jouV2V1}.
As in the case of infinitely thin airfoils, these DDLSs are also impressively similar to the PLS shown in figure \ref{fig.clabeta0}. Again they look  rotations, reflections or stretching of the same curve.

\begin{figure}
    \centering

   \begin{tikzpicture}
    \pgfplotsset{every axis legend/.append style={at={(1,0)},anchor=south east}}
    \begin{axis}[height=8cm, width=6cm, grid, xmin=0, xmax=1,ymin=0,ymax=1,xlabel=$\hat{\alpha}$, ylabel=$\hat{C}_l$]
    \addplot[ultra thick,solid, color=cyan, mark size = 3pt]
    table [x expr=\thisrowno{0}, y expr=\thisrowno{1}] {plots/J_PLS_affine_normalized.txt};
    \addplot[solid, color=magenta, mark size = 3pt]
    table [x expr=\thisrowno{0}, y expr=\thisrowno{1}] {plots/J_training1_affine_normalized.txt};
     \addplot[solid, color=green, mark size = 3pt]
    table [x expr=\thisrowno{0}, y expr=\thisrowno{1}] {plots/J_training2_affine_normalized.txt};
     \addplot[solid, color=blue, mark size = 3pt]
    table [x expr=\thisrowno{0}, y expr=\thisrowno{1}] {plots/J_training3_affine_normalized.txt};
     \addplot[solid, color=orange, mark size = 3pt]
    table [x expr=\thisrowno{0}, y expr=\thisrowno{1}] {plots/J_training4_affine_normalized.txt};
    \legend{\tiny PLS, \tiny training (a), \tiny training (b), \tiny training (c), \tiny training (d)
 }
    \end{axis}
    \end{tikzpicture}

    \caption{Joukowski airfoil $R=1.1$, $\beta=0$. Comparison of normalized $(\alpha, C_l)$ PLS with normalized DDLS obtained in 4 trainings of a vanilla AE after GAT.}
    \label{fig.jouv1v2norm}
\end{figure}
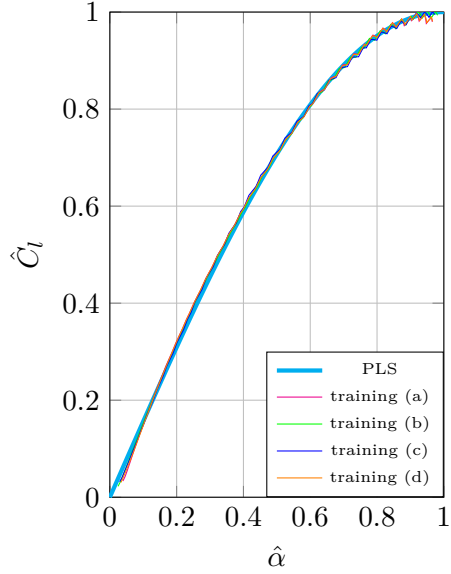

Table \ref{tab.jou1} shows the coefficients of determination $R^2$  of the ridge regression between the 4 DDLSs and the parameters AoA and lift coefficient.  The obtained extremely high $R^2$ values, also in this case, reveal the very strong correlation between the PLS and all obtained DDLSs:  we had all $R^2>0.99$.
\begin{table}[]
    \centering
    \begin{tabular}{|c|c|c|c|c|c|c|}
    \hline
    \hline
    DDLS& $\alpha$&$C_l$& $E_{P}$ & $E_{A}$\\
  \hline
  \hline
  Training (a) &0.9971&0.9987&0.1562 & 0.0501\\
 Training  (b) &0.9992&0.9995&0.1488& 0.0267\\
 Training  (c) &0.9984&0.9992&0.0955& 0.0383\\
 Training  (d) &0.9977&0.9988&0.1380 & 0.0460 \\
  \hline
  \hline
    \end{tabular}
    \caption{Joukowski airfoil $R=1.1$, $\beta=0$. 
    Results for 4 trainings. 
    $R^2$ values of ridge regression
    on AoA and lift coefficient 
    and normalized errors of PA and GAT of the DDLS compared with PLS ($\alpha,C_l$). }
    \label{tab.jou1}
\end{table}
Table \ref{tab.jou1} also shows the normalized PA and GAT errors. For the 4 independent AE trainings, the normalized PA error ranges from 0.0955 to 0.1562, whereas the normalized error obtained after a GAT ranges from 0.0267 to 0.0501. The non-overlapping intervals indicate that, in all trainings, the DDLS is again substantially better approximated as an affine deformation of the PLS than as a similarity transformation.
After applying the independently determined affine transformation to each training, the four DDLS realizations collapse closely onto the normalized PLS as evidenced by figure \ref{fig.jouv1v2norm}. The residual deviations are small and are mainly confined to the ends of the curve, confirming that the different AE realizations recover approximately the same two-dimensional physical manifold up to an affine transformation.
The small oscillations, more evident at larger AoAs, are a numerical artifact due to the discrete sampling of the continuous Joukowski field.
This result demonstrates that a two-dimensional DDLS embeds the one-dimensional physical trajectory in a geometry closely related to the physically constructed  $(\alpha, C_l)$ embedding.
The analysis was made possible by the relatively simple cases 
we studied which were ruled by one or maximum two physical parameters. 
In more complex problems, due to the complexity of the latent space manifold, the physical relevance of the latent variables could be hidden, but still present.

\section{Case 3: Airfoils at high Reynolds number }
\label{NACA}

The aerodynamic database studied in this section consists of  flow fields around airfoils at high Reynolds numbers. Turbulent flow at Mach number $M_\infty=0.15$ and Reynolds number $Re_\infty=5 \times 10^6$ was assumed. The database (here named NACA db 1) is built by 124 CFD calculations around four NACA airfoils (0004, 0006, 0008, 0010) \cite{Abbott1959TheoryOW}, with  $\alpha \in [0^\circ,30^\circ]$, step $1^\circ$. The chosen airfoils are symmetric with increasing maximum thickness $t=\{4\%, 6\%, 8\%, 10\% \}$. 
The numerical solutions were generated by the RANS open source solver SU2, originally developed at Stanford University \cite{Economon2016SU2} adopting Spalart-Allmaras turbulence model. 
The grids are O-type with 512 cells around the airfoil and 256 cells in normal direction.  
The distance of the far-field boundary from the body is about 50 airfoil chords. Each data point has 5 channels, pressure coefficient, two cartesian velocity components and $x-y$ grid coordinates. Since in the far field the flow is essentially uniform, we preferred to focus the training in the region around the airfoil; therefore, we reduced the grid stored in the database to $257 \times 191$ points with the flow field extending to a distance of approximately one airfoil chord from the body.

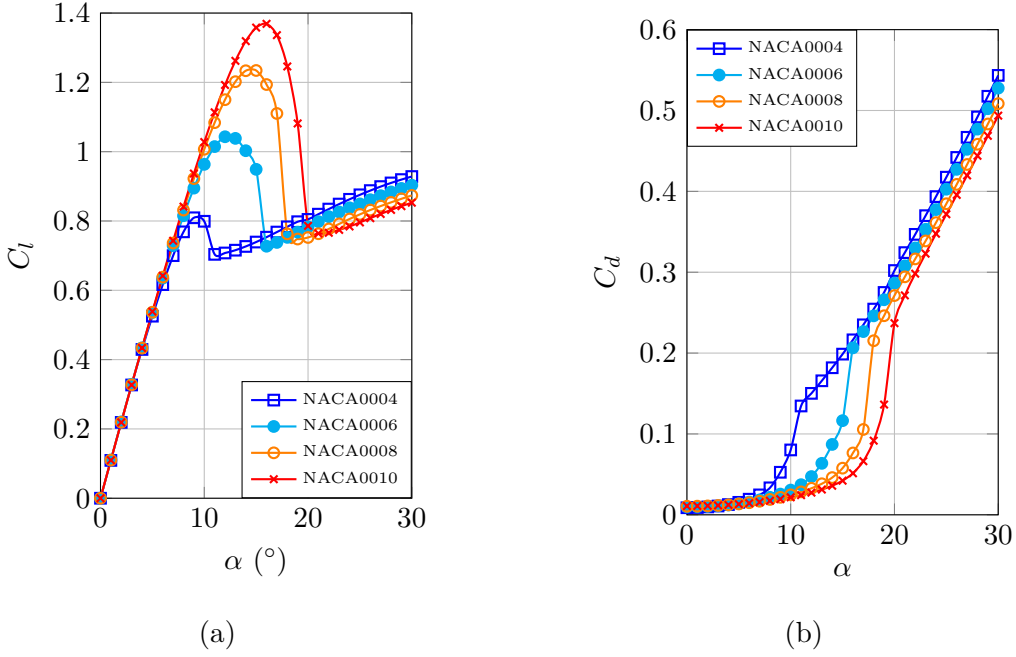
\begin{figure*}
\centering
\begin{subfigure}{0.45\textwidth}
\centering
\begin{tikzpicture}
\pgfplotsset{every axis legend/.append style={at={(1,0)},anchor=south east}}
\begin{axis}[height=8.0cm,width=5.7cm,
grid,xlabel=$\alpha$  $(^\circ)$, ylabel=$C_l$, xmin=0, xmax=30,ymin=0,ymax=1.4]
\addplot[thick,blue,solid,smooth,mark=square]
 table [x expr=\thisrowno{0}, y expr=\thisrowno{1}] {plots/AeroForces_n0004.dat}; 
\addplot[thick,cyan,solid,smooth,mark=*]
 table [x expr=\thisrowno{0}, y expr=\thisrowno{1}] 
 {plots/AeroForces_n0006.dat}; 
 \addplot[thick,orange,solid,smooth,mark=o]
 table [x expr=\thisrowno{0}, y expr=\thisrowno{1}] 
  {plots/AeroForces_n0008.dat}; 
 \addplot[thick,red,solid,smooth,mark=x]
 table [x expr=\thisrowno{0}, y expr=\thisrowno{1}] 
  {plots/AeroForces_n0010.dat}; 
 \legend{\tiny NACA0004, \tiny NACA0006, \tiny NACA0008, \tiny NACA0010
 }
\end{axis}
\end{tikzpicture}
\caption{}
\end{subfigure}
\hspace{1mm}
\begin{subfigure}{0.45\textwidth}
\centering
\begin{tikzpicture}
\pgfplotsset{every axis legend/.append style={at={(0,1)},anchor=north west}}
\begin{axis}[height=8.0cm,width=5.7cm,
grid,xlabel=$\alpha$, ylabel=$C_d$, xmin=0, xmax=30,ymin=0,ymax=0.6]
\addplot[thick,blue,solid,smooth,mark=square]
 table [x expr=\thisrowno{0}, y expr=\thisrowno{2}] {plots/AeroForces_n0004.dat}; 
\addplot[thick,cyan,solid,smooth,mark=*]
 table [x expr=\thisrowno{0}, y expr=\thisrowno{2}] 
 {plots/AeroForces_n0006.dat}; 
 \addplot[thick,orange,solid,smooth,mark=o]
 table [x expr=\thisrowno{0}, y expr=\thisrowno{2}] 
  {plots/AeroForces_n0008.dat}; 
 \addplot[thick,red,solid,smooth,mark=x]
 table [x expr=\thisrowno{0}, y expr=\thisrowno{2}] 
   {plots/AeroForces_n0010.dat}; 
 \legend{\tiny NACA0004, \tiny NACA0006, \tiny NACA0008,\tiny NACA0010}
\end{axis}
\end{tikzpicture}
\caption{}
\end{subfigure}
    \caption{NACA db 1. Lift (a) and drag coefficient (b) versus AoA.  
    $M_\infty=0.15$, $Re_\infty=5 \times 10^6$.}
    \label{fig.data1}
\end{figure*}

Thin airfoils with a wide range of AoAs  were chosen in order to increase the complexity of the flows. Indeed, as shown in figure \ref{fig.data1} where the lift and drag coefficients are plotted
against AoA, 
very strong non linearities are present, all airfoils encounter 4 different regimes: linear increase of lift, stall, post-stall and massively separated flow with lift recovery at the highest AoAs.
In addition, the maximum lift coefficients present large variations with thickness, from 0.8 to 1.37.
Even if the engineering accuracy of the RANS simulations could be poor for the highest AoAs, they qualitatively reproduce the correct physics and are assumed as ground truth.

\begin{figure*}
\centering
\begin{subfigure}{0.45\textwidth}
\begin{tikzpicture}
\begin{axis}[width=6cm,height=6.0cm,
grid,xlabel=$V_1$, ylabel=$V_2$]
\addplot[only marks,magenta,mark=*]
 table [x expr=\thisrowno{2}, y expr=\thisrowno{3}] {plots/NACA_alpha_V1_V2_a.txt}; 
 \draw[->, thick]
    (axis cs:-0.2,0.1) --
    (axis cs:-0.1,0.2)
    node[midway, above, sloped] {$\alpha$};
 \draw[->, thick]
    (axis cs:-0.25,0.45) --
    (axis cs:-0.35,0.55)
    node[midway, above, sloped] {$t$};
\end{axis}
\end{tikzpicture}
\caption{}
\end{subfigure}
\hspace{5mm}
\begin{subfigure}{0.45\textwidth}
\begin{tikzpicture}
\begin{axis}[width=6cm,height=6.0cm,
grid,xlabel=$V_1$, ylabel=$V_2$]
\addplot[only marks,green,mark=*]
 table [x expr=\thisrowno{2}, y expr=\thisrowno{3}] {plots/NACA_alpha_V1_V2_f.txt}; 
 \draw[->, thick]
    (axis cs:0.5,0.1) --
    (axis cs:0.35,0.05)
    node[midway, above, sloped] {$\alpha$};
 \draw[->, thick]
    (axis cs:0.3,-0.10) --
    (axis cs:0.4,-0.2)
    node[midway, above, sloped] {$t$};
\end{axis}
\end{tikzpicture}
\caption{}
\end{subfigure}

\vspace{5mm}
\begin{subfigure}{0.45\textwidth}
\begin{tikzpicture}
\begin{axis}[width=6cm,height=6.0cm,
grid,xlabel=$V_1$, ylabel=$V_2$]
\addplot[only marks,blue,mark=*]
 table [x expr=\thisrowno{2}, y expr=\thisrowno{3}] {plots/NACA_alpha_V1_V2_c.txt}; 
 \draw[->, thick]
    (axis cs:-0.2,0.2) --
    (axis cs:-0.3,0.0)
    node[midway, above, sloped] {$\alpha$};
 \draw[->, thick]
    (axis cs:-0.6,0.0) --
    (axis cs:-0.8,0.2)
    node[midway, above, sloped] {$t$};
\end{axis}
\end{tikzpicture}
\caption{}
\end{subfigure}
\hspace{5mm}
\begin{subfigure}{0.45\textwidth}
\begin{tikzpicture}
\begin{axis}[width=6cm,height=6.0cm,
grid,xlabel=$V_1$, ylabel=$V_2$]
\addplot[only marks,orange,mark=*]
 table [x expr=\thisrowno{2}, y expr=\thisrowno{3}] {plots/NACA_alpha_V1_V2_d.txt};
 \draw[->, thick]
    (axis cs:-0.2,-0.2) --
    (axis cs:-0.2,0.0)
    node[midway, above, sloped] {$\alpha$};
 \draw[->, thick]
    (axis cs:-0.4,0.0) --
    (axis cs:-0.55,0.0)
    node[midway, above, sloped] {$t$};
\end{axis}
\end{tikzpicture}
\caption{}
\end{subfigure}
\caption{NACA db 1.
DDLS as obtained by 4 different trainings of vanilla AE with $P=2$.}
    \label{fig.nacaaetrains}
\end{figure*}
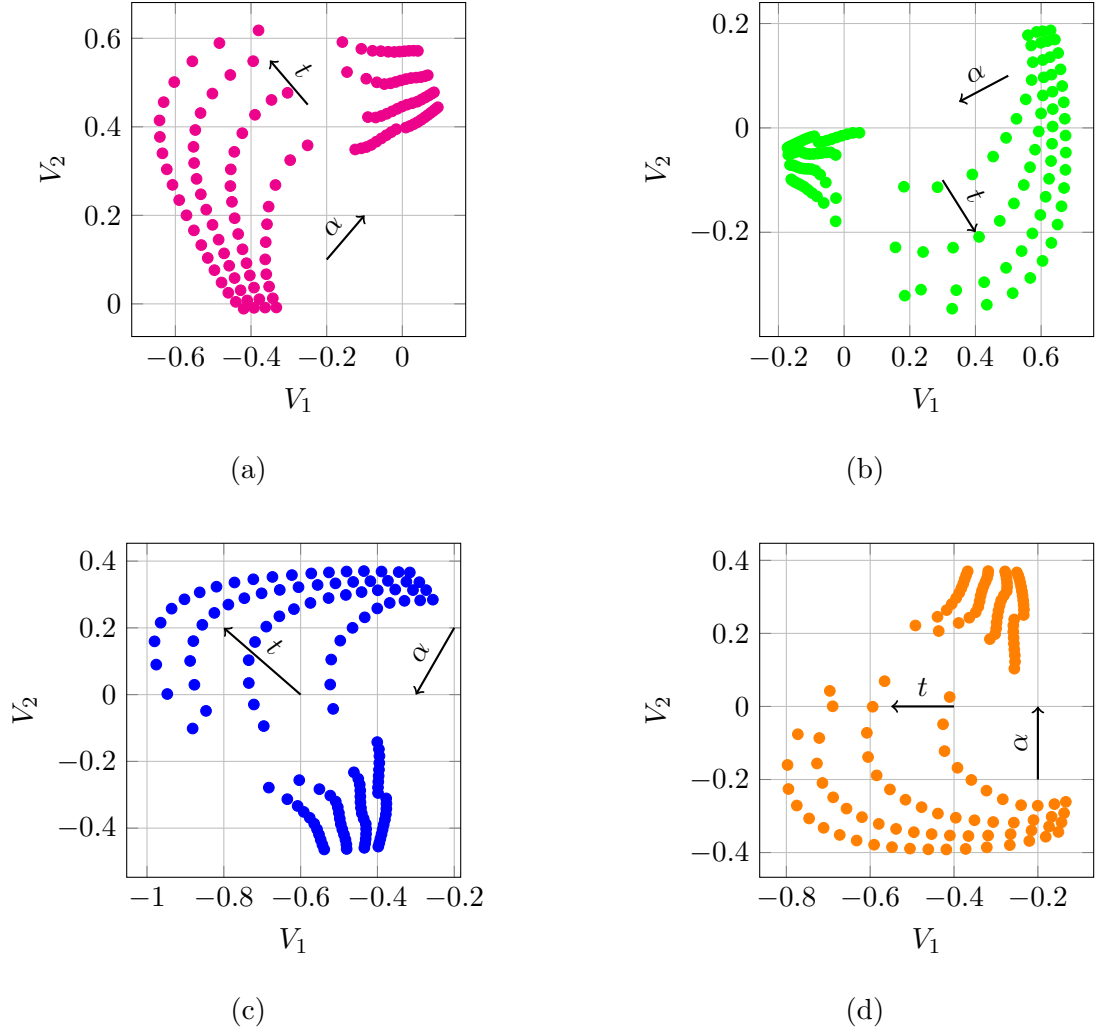

Both figures \ref{fig.data1}(a) and \ref{fig.data1}(b) are   two  alternative PLS of the database. In the first one, the latent variables are lift  and AoA
parametrized with airfoil thickness; in the second one drag and AoA, again 
parametrized with  thickness.
Figure \ref{fig.nacaaetrains}
shows 4 random trainings adopting the vanilla AE and latent space dimension $P=2$.
Even if they are rotated, deformed or reflected, the shape of all obtained DDLS are undoubtedly very similar to both PLS (lift vs AoA and drag against AoA) and again show a clear physical imprinting..  
The linear regime, stall, post-stall and lift recovery regions are clearly recognizable in the DDLS. 
The effect of thickness $t$ on maximum lift and stall angle is also correctly reproduced and the maximum lift coefficient point is obtained at the correct corresponding AoA.

Table \ref{tab.naca1} analyses  the correlation of the 4 latent spaces of figure  \ref{fig.nacaaetrains} with the physical parameters $\alpha$ and $t$ and the physical responses $C_l$ and $C_d$ by ridge regression.
\begin{table}[]
    \centering
    \begin{tabular}{|c|c|c|c|c|}
    \hline
    \hline
    DDLS& $\alpha$&$t$&$C_l$&$C_d$\\
  \hline
  \hline
  Training (a) &0.909&0.288&0.773&0.891\\
  Training (b) &0.876&0.128&0.838&0.864\\
  Training (c) &0.935&0.279&0.827&0.903\\
  Training (d) &0.908&0.200&0.842&0.886\\
  \hline
  \hline
    \end{tabular}
    \caption{NACA db 1. 
    $R^2$ values of ridge regression
    on physical parameters and responses
    of 4 trained DDLS.}
    \label{tab.naca1}
\end{table}
\begin{table}[]
    \centering
    \begin{tabular}{|c|c|c|c|c|}
    \hline
    \hline
     \multicolumn{1}{|c|}{}
     &\multicolumn{2}{|c|}{PLS ($\alpha,C_l$)} 
     &\multicolumn{2}{|c|}{PLS ($\alpha,C_d$)} \\
     \hline
     DDLS &$E_{P}$&$E_{A}$ &$E_{P}$ &$E_{A}$ \\ 
     \hline
     \hline
     Training (a) &0.6832&0.4229&0.4058&0.3018\\
     Training (b) &0.7367&0.4153&0.4072&0.3689\\
     Training (c) &0.6876&0.4434&0.4567&0.3044\\
     Training (d) &0.6788&0.3504&0.4523&0.3093\\
  \hline
  \hline
    \end{tabular}
    \caption{NACA db 1. 
    Normalized errors of PA and GAT of the DDLS compared with PLS ($\alpha,C_l$) and PLS ($\alpha,C_d$).}
    \label{tab.naca1PA}
\end{table}
It shows that all DDLS are very well correlated with the $(\alpha, C_d)$ PLS; the correlation with the the lift coefficient is present but does not fulfill the requirement $R^2 > 0.85$. The correlation with thickness is quite low; this result is interesting because  again the AE,
instead of  organizing its representation according to the parameters used to generate the dataset,   organized it according to the \emph{resulting} physical state or aerodynamic response.
However, we should note that part of the correlation with $C_d$ may be inherited from $\alpha$, and that thickness induces comparatively small changes in the flow field. The dominant latent directions follow the largest variations of the flow state, which in this case are better described by aerodynamic responses than by the thickness parameter.

Table \ref{tab.naca1PA} proposes the PA and GAT of the 4 trained DDLS against the PLS $(\alpha,C_l)$ and $(\alpha,C_d)$. This time the error is very high, with the affine transformation  always better. 
However this result needs further investigation.
Indeed a high $R^2$ ridge regression value implies, for instance
\begin{equation}
    \alpha \approx b_0+ b_1 V_1+ b_2 V_2 \ .
\end{equation}
So $R^2>0.85$ means that $\alpha$ is strongly encoded as a linear combination of the two latent variables. It does not require the complete geometry of the DDLS to look like the corresponding physical latent space.
However there is a close relation with the general affine transformation because mapping
\begin{equation}
     (V_1,V_2)\rightarrow(\alpha,C_d) 
\end{equation}
is equivalent to perform two simultaneous linear regressions, one for $\alpha$ and one for $C_d$.
Since  both physical coordinates are standardized to variance 1, for an ordinary least-squares affine fit,
\begin{equation}
      E_{A}^2 = 1-\frac{(R_\alpha^2+R_{C_d}^2)}{2} 
\end{equation}
 where  $R_\alpha^2$   and $R_{C_d}^2$ are respectively the coefficients of determination of the ridge regression computed using $\alpha$ and $C_d$. Therefore the average coefficient of determination  $\overline{R^2}$ is
\begin{equation}
  \overline{R^2}=1-E_{A}^2  \ ,
\end{equation}
 result strictly valid in case of ordinary least squares on the complete dataset and not for cross-correlation.
Taking for instance the value $E_{A}=0.3018$ of the first training against PLS $(\alpha, C_d)$ we have $\overline{R^2} =0.909$ that indeed is approximately the same result of the ridge regression analysis. The result could be therefore better  reformulated as: the affine result tells that about 91\% of the combined variance of $\alpha$ and $C_d$ can be reproduced linearly from $(V_1,V_2)$.
the DDLS does not visually or geometrically reproduce the PLS under a similarity transformation, but the physical variables are nevertheless strongly and linearly encoded in it. The affine analysis provides the bridge between those two observations.

\section{Exploiting physical latent structure: separation of Reynolds-number effects}

The database representing the ground truth (here named NACA db 2) is built with 272 solutions around the symmetric NACA 0012, 0014, 0016 and 0018 airfoils (thickness $t=\{12\%, 14\%, 16\%, 18\%\}$), at Reynolds number $Re_\infty = \{10^5, 10^6, 5\times 10^6, \infty\}$,  $\alpha \in [0^\circ,16^\circ]$ step $1^\circ$ and $M_\infty=0.15$.
The inviscid cases ($Re_\infty=\infty$) are again obtained by SU2 solver, this time solving the Euler equations. The size of the grids is the same as in previous section.
The PLSs, lift against AoA and drag against AoA, are plotted in figures \ref{fig.nReCl} and
\ref{fig.nReCd}, separating for clarity each airfoil.
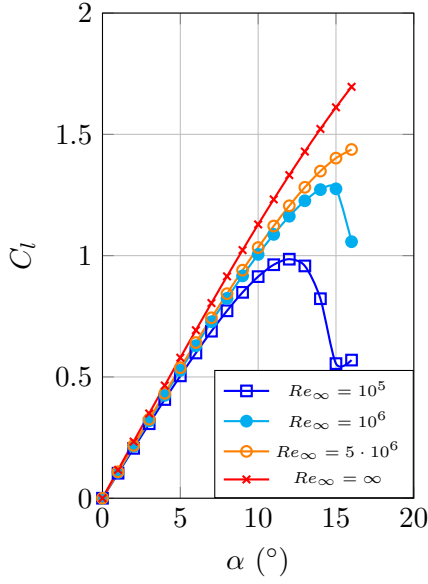
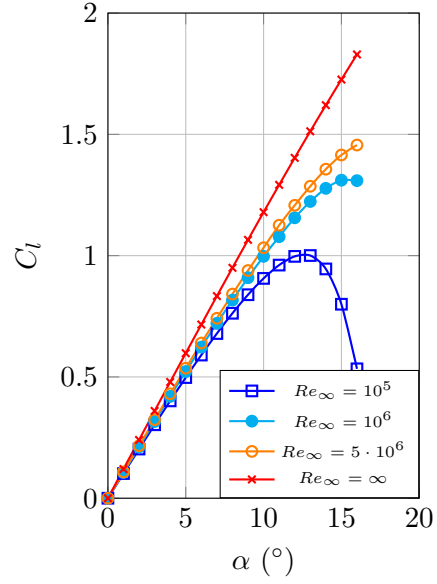
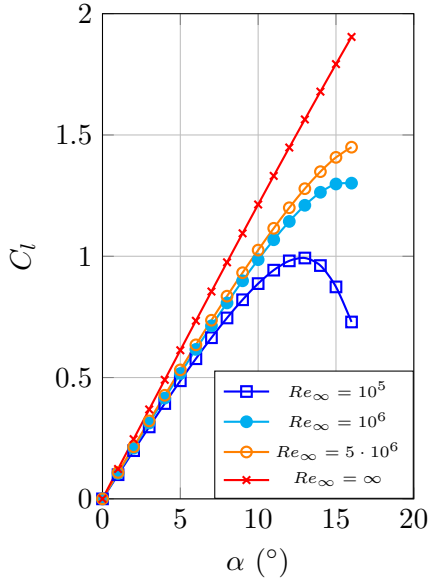
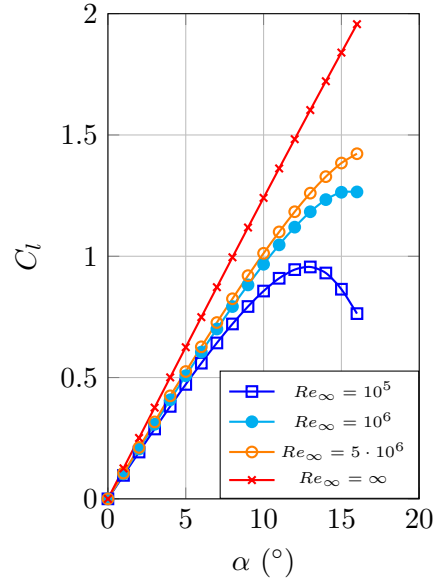
\begin{figure*}
\centering
\begin{subfigure}{0.45\textwidth}
\centering
\begin{tikzpicture}
\pgfplotsset{every axis legend/.append style={at={(1,0)},anchor=south east}}
\begin{axis}[height=8.0cm,width=5.7cm,
grid,xlabel=$\alpha$  $(^\circ)$, ylabel=$C_l$, xmin=0, xmax=20,ymin=0,ymax=2.0]
\addplot[thick,blue,solid,smooth,mark=square]
 table [x expr=\thisrowno{0}, y expr=\thisrowno{1}] {plots/N0012Re.dat}; 
\addplot[thick,cyan,solid,smooth,mark=*]
 table [x expr=\thisrowno{0}, y expr=\thisrowno{3}] 
 {plots/N0012Re.dat}; 
 \addplot[thick,orange,solid,smooth,mark=o]
 table [x expr=\thisrowno{0}, y expr=\thisrowno{5}] 
  {plots/N0012Re.dat}; 
 \addplot[thick,red,solid,smooth,mark=x]
 table [x expr=\thisrowno{0}, y expr=\thisrowno{7}] 
  {plots/N0012Re.dat}; 
 \legend{\tiny $Re_\infty=10^5$, \tiny $Re_\infty=10^6$, \tiny $Re_\infty=5 \cdot 10^6$, \tiny $Re_\infty=\infty$
 }
\end{axis}
\end{tikzpicture}
\caption{NACA0012}
\end{subfigure}
\hspace{1mm}
\begin{subfigure}{0.45\textwidth}
\centering
\begin{tikzpicture}
\pgfplotsset{every axis legend/.append style={at={(1,0)},anchor=south east}}
\begin{axis}[height=8.0cm,width=5.7cm,
grid,xlabel=$\alpha$  $(^\circ)$, ylabel=$C_l$, xmin=0, xmax=20,ymin=0,ymax=2.0]
\addplot[thick,blue,solid,smooth,mark=square]
 table [x expr=\thisrowno{0}, y expr=\thisrowno{1}] {plots/N0014Re.dat}; 
\addplot[thick,cyan,solid,smooth,mark=*]
 table [x expr=\thisrowno{0}, y expr=\thisrowno{3}] 
 {plots/N0014Re.dat}; 
 \addplot[thick,orange,solid,smooth,mark=o]
 table [x expr=\thisrowno{0}, y expr=\thisrowno{5}] 
  {plots/N0014Re.dat}; 
 \addplot[thick,red,solid,smooth,mark=x]
 table [x expr=\thisrowno{0}, y expr=\thisrowno{7}] 
  {plots/N0014Re.dat}; 
 \legend{\tiny $Re_\infty=10^5$, \tiny $Re_\infty=10^6$, \tiny $Re_\infty=5 \cdot 10^6$, \tiny $Re_\infty=\infty$
 }
\end{axis}
\end{tikzpicture}
\caption{NACA0014}
\end{subfigure}

\begin{subfigure}{0.45\textwidth}
\centering
\begin{tikzpicture}
\pgfplotsset{every axis legend/.append style={at={(1,0)},anchor=south east}}
\begin{axis}[height=8.0cm,width=5.7cm,
grid,xlabel=$\alpha$  $(^\circ)$, ylabel=$C_l$, xmin=0, xmax=20,ymin=0,ymax=2.0]
\addplot[thick,blue,solid,smooth,mark=square]
 table [x expr=\thisrowno{0}, y expr=\thisrowno{1}] {plots/N0016Re.dat}; 
\addplot[thick,cyan,solid,smooth,mark=*]
 table [x expr=\thisrowno{0}, y expr=\thisrowno{3}] 
 {plots/N0016Re.dat}; 
 \addplot[thick,orange,solid,smooth,mark=o]
 table [x expr=\thisrowno{0}, y expr=\thisrowno{5}] 
  {plots/N0016Re.dat}; 
 \addplot[thick,red,solid,smooth,mark=x]
 table [x expr=\thisrowno{0}, y expr=\thisrowno{7}] 
  {plots/N0016Re.dat}; 
 \legend{\tiny $Re_\infty=10^5$, \tiny $Re_\infty=10^6$, \tiny $Re_\infty=5 \cdot 10^6$, \tiny $Re_\infty=\infty$
 }
\end{axis}
\end{tikzpicture}
\caption{NACA0016}
\end{subfigure}
\hspace{1mm}
\begin{subfigure}{0.45\textwidth}
\centering
\begin{tikzpicture}
\pgfplotsset{every axis legend/.append style={at={(1,0)},anchor=south east}}
\begin{axis}[height=8.0cm,width=5.7cm,
grid,xlabel=$\alpha$  $(^\circ)$, ylabel=$C_l$, xmin=0, xmax=20,ymin=0,ymax=2.0]
\addplot[thick,blue,solid,smooth,mark=square]
 table [x expr=\thisrowno{0}, y expr=\thisrowno{1}] {plots/N0018Re.dat}; 
\addplot[thick,cyan,solid,smooth,mark=*]
 table [x expr=\thisrowno{0}, y expr=\thisrowno{3}] 
 {plots/N0018Re.dat}; 
 \addplot[thick,orange,solid,smooth,mark=o]
 table [x expr=\thisrowno{0}, y expr=\thisrowno{5}] 
  {plots/N0018Re.dat}; 
 \addplot[thick,red,solid,smooth,mark=x]
 table [x expr=\thisrowno{0}, y expr=\thisrowno{7}] 
  {plots/N0018Re.dat}; 
 \legend{\tiny $Re_\infty=10^5$, \tiny $Re_\infty=10^6$, \tiny $Re_\infty=5 \cdot 10^6$, \tiny $Re_\infty=\infty$
 }
\end{axis}
\end{tikzpicture}
\caption{NACA0018}
\end{subfigure}

\caption{NACA db 2. $M_\infty=0.15$.
Lift coefficient versus AoA.}
\label{fig.nReCl}

\end{figure*}
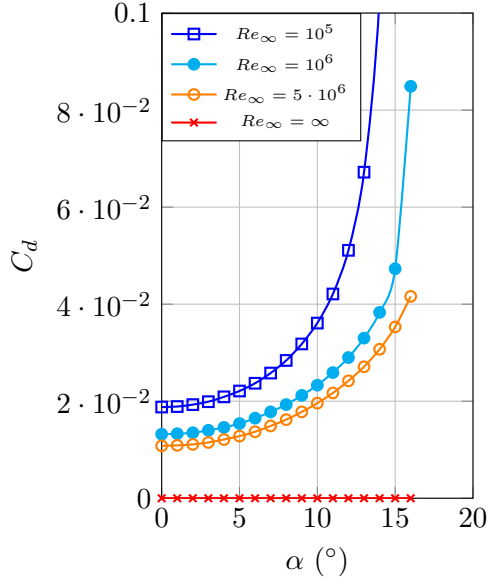
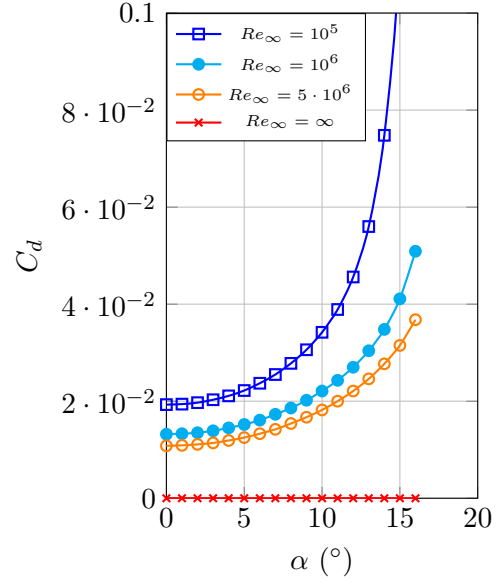
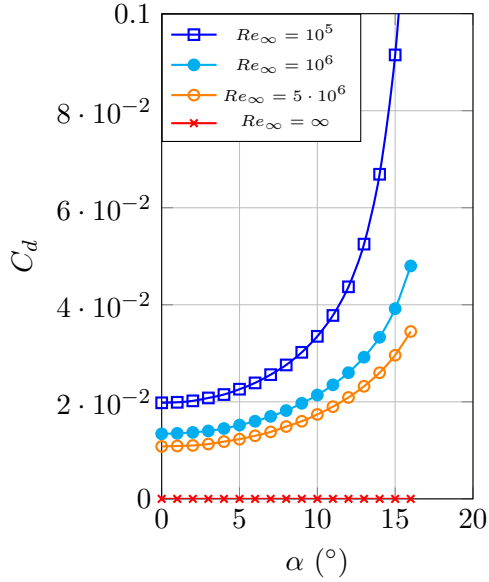
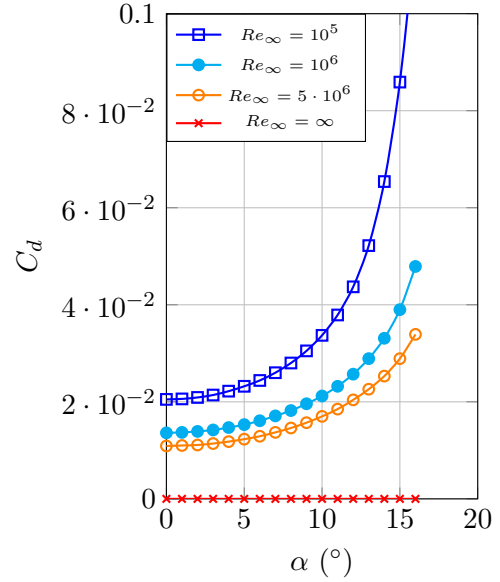
\begin{figure*}
\centering
\begin{subfigure}{0.48\textwidth}
\centering
\begin{tikzpicture}
\pgfplotsset{every axis legend/.append style={at={(0,1)},anchor=north west}}
\begin{axis}[height=8.0cm,width=5.7cm,
grid,xlabel=$\alpha$  $(^\circ)$, ylabel=$C_d$, xmin=0, xmax=20,ymin=0,ymax=0.1]
\addplot[thick,blue,solid,smooth,mark=square]
 table [x expr=\thisrowno{0}, y expr=\thisrowno{2}] {plots/N0012Re.dat}; 
\addplot[thick,cyan,solid,smooth,mark=*]
 table [x expr=\thisrowno{0}, y expr=\thisrowno{4}] 
 {plots/N0012Re.dat}; 
 \addplot[thick,orange,solid,smooth,mark=o]
 table [x expr=\thisrowno{0}, y expr=\thisrowno{6}] 
  {plots/N0012Re.dat}; 
 \addplot[thick,red,solid,smooth,mark=x]
 table [x expr=\thisrowno{0}, y expr=0] 
  {plots/N0012Re.dat}; 
 \legend{\tiny $Re_\infty=10^5$, \tiny $Re_\infty=10^6$, \tiny $Re_\infty=5 \cdot 10^6$, \tiny $Re_\infty=\infty$
 }
\end{axis}
\end{tikzpicture}
\caption{NACA0012}
\end{subfigure}
\hspace{1mm}
\begin{subfigure}{0.48\textwidth}
\centering
\begin{tikzpicture}
\pgfplotsset{every axis legend/.append style={at={(0,1)},anchor=north west}}
\begin{axis}[height=8.0cm,width=5.7cm,
grid,xlabel=$\alpha$  $(^\circ)$, ylabel=$C_d$, xmin=0, xmax=20,ymin=0,ymax=0.1]
\addplot[thick,blue,solid,smooth,mark=square]
 table [x expr=\thisrowno{0}, y expr=\thisrowno{2}] {plots/N0014Re.dat}; 
\addplot[thick,cyan,solid,smooth,mark=*]
 table [x expr=\thisrowno{0}, y expr=\thisrowno{4}] 
 {plots/N0014Re.dat}; 
 \addplot[thick,orange,solid,smooth,mark=o]
 table [x expr=\thisrowno{0}, y expr=\thisrowno{6}] 
  {plots/N0014Re.dat}; 
 \addplot[thick,red,solid,smooth,mark=x]
 table [x expr=\thisrowno{0}, y expr=0] 
  {plots/N0014Re.dat}; 
 \legend{\tiny $Re_\infty=10^5$, \tiny $Re_\infty=10^6$, \tiny $Re_\infty=5 \cdot 10^6$, \tiny $Re_\infty=\infty$
 }
\end{axis}
\end{tikzpicture}
\caption{NACA0014}
\end{subfigure}

\begin{subfigure}{0.48\textwidth}
\centering
\begin{tikzpicture}
\pgfplotsset{every axis legend/.append style={at={(0,1)},anchor=north west}}
\begin{axis}[height=8.0cm,width=5.7cm,
grid,xlabel=$\alpha$  $(^\circ)$, ylabel=$C_d$, xmin=0, xmax=20,ymin=0,ymax=0.1]
\addplot[thick,blue,solid,smooth,mark=square]
 table [x expr=\thisrowno{0}, y expr=\thisrowno{2}] {plots/N0016Re.dat}; 
\addplot[thick,cyan,solid,smooth,mark=*]
 table [x expr=\thisrowno{0}, y expr=\thisrowno{4}] 
 {plots/N0016Re.dat}; 
 \addplot[thick,orange,solid,smooth,mark=o]
 table [x expr=\thisrowno{0}, y expr=\thisrowno{6}] 
  {plots/N0016Re.dat}; 
 \addplot[thick,red,solid,smooth,mark=x]
 table [x expr=\thisrowno{0}, y expr= 0] 
  {plots/N0016Re.dat}; 
 \legend{\tiny $Re_\infty=10^5$, \tiny $Re_\infty=10^6$, \tiny $Re_\infty=5 \cdot 10^6$, \tiny $Re_\infty=\infty$
 }
\end{axis}
\end{tikzpicture}
\caption{NACA0016}
\end{subfigure}
\hspace{1mm}
\begin{subfigure}{0.48\textwidth}
\centering
\begin{tikzpicture}
\pgfplotsset{every axis legend/.append style={at={(0,1)},anchor=north west}}
\begin{axis}[height=8.0cm,width=5.7cm,
grid,xlabel=$\alpha$  $(^\circ)$, ylabel=$C_d$, xmin=0, xmax=20,ymin=0,ymax=0.1]
\addplot[thick,blue,solid,smooth,mark=square]
 table [x expr=\thisrowno{0}, y expr=\thisrowno{2}] {plots/N0018Re.dat}; 
\addplot[thick,cyan,solid,smooth,mark=*]
 table [x expr=\thisrowno{0}, y expr=\thisrowno{4}] 
 {plots/N0018Re.dat}; 
 \addplot[thick,orange,solid,smooth,mark=o]
 table [x expr=\thisrowno{0}, y expr=\thisrowno{6}] 
  {plots/N0018Re.dat}; 
 \addplot[thick,red,solid,smooth,mark=x]
 table [x expr=\thisrowno{0}, y expr=0] 
  {plots/N0018Re.dat}; 
 \legend{\tiny $Re_\infty=10^5$, \tiny $Re_\infty=10^6$, \tiny $Re_\infty=5 \cdot 10^6$, \tiny $Re_\infty=\infty$
 }
\end{axis}
\end{tikzpicture}
\caption{NACA0018}
\end{subfigure}

\caption{NACA db 2. $M_\infty=0.15$.
Drag coefficient versus AoA.}
\label{fig.nReCd}

\end{figure*}

The training is  performed adopting the multi-fidelity AESTAR   described in section \ref{sec.AE}.
During the first step, the training is performed only analyzing the inviscid data using the vanilla AE with latent space dimension $p=2$ while the third latent variable is assumed zero. 
The second step consists in training with $P=3$ both inviscid  and viscous data, but forcing, with a proper loss function, the first 2 latent variables to be equal to the ones identified in the previous step.
The third latent variable $V_3$ is free to be learnt and is expected to identify the Reynolds number effects, with $V_3=0$ indicating inviscid  flow.

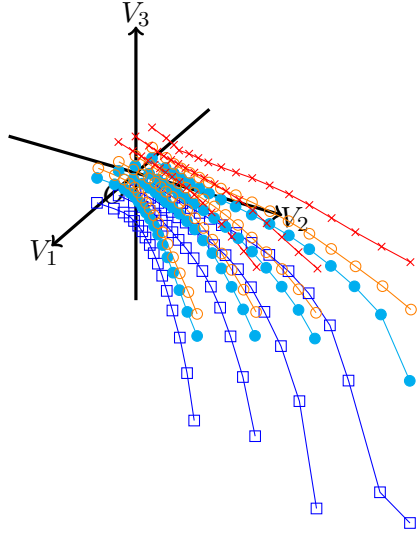
\begin{figure}
\centering
\begin{tikzpicture}
\begin{axis}
[ width=10cm, height=10cm, 
view={120}{30}, 
xmin=-1, xmax=1, 
ymin=-0.5, ymax=2.5, 
zmin=-2.5, zmax=0.001, 
axis lines=none, 
axis equal image, 
xtick=\empty, ytick=\empty, ztick=\empty, 
clip=false 
]
\addplot3[ ->, very thick, black ] coordinates { (-1,0,0) (1.15,0,0) };
\addplot3[ ->, very thick, black ] coordinates { (0,-1,0) (0,1.15,0) }; 
\addplot3[ ->, very thick, black ] coordinates { (0,0,-1) (0,0,1.15) }; 
\node at (axis cs:1.25,0,0) {$V_1$}; 
\node at (axis cs:0,1.25,0) {$V_2$}; 
\node at (axis cs:0,0,1.25) {$V_3$}; 
\node[ anchor=north east ] at (axis cs:0,0,0) {$O$};

\addplot3[blue,mark=square,
     x filter/.code={
        \ifnum\coordindex<0
             \def\pgfmathresult{}
         \fi
         \ifnum\coordindex>16
             \def\pgfmathresult{}
         \fi
     },
]
 table [x index = 2,
        y index = 3,
        z index = 4
 ]
{plots/NACA_ReEffect_V1_V2_V3_old.txt}; 

\addplot3[blue,mark=square,
     x filter/.code={
        \ifnum\coordindex<51
             \def\pgfmathresult{}
         \fi
         \ifnum\coordindex>67
             \def\pgfmathresult{}
         \fi
     },
]
 table [x index = 2,
        y index = 3,
        z index = 4
 ]
{plots/NACA_ReEffect_V1_V2_V3_old.txt}; 
\addplot3[blue,mark=square,
     x filter/.code={
        \ifnum\coordindex<102
             \def\pgfmathresult{}
         \fi
         \ifnum\coordindex>118
             \def\pgfmathresult{}
         \fi
     },
]
 table [x index = 2,
        y index = 3,
        z index = 4
 ]
{plots/NACA_ReEffect_V1_V2_V3_old.txt}; 
\addplot3[blue,mark=square,
     x filter/.code={
        \ifnum\coordindex<153
             \def\pgfmathresult{}
         \fi
         \ifnum\coordindex>169
             \def\pgfmathresult{}
         \fi
     },
]
 table [x index = 2,
        y index = 3,
        z index = 4
 ]
{plots/NACA_ReEffect_V1_V2_V3_old.txt}; 

\addplot3[cyan,mark=*,
     x filter/.code={
        \ifnum\coordindex<17
             \def\pgfmathresult{}
         \fi
         \ifnum\coordindex>33
             \def\pgfmathresult{}
         \fi
     },
]
 table [x index = 2,
        y index = 3,
        z index = 4
 ]
{plots/NACA_ReEffect_V1_V2_V3_old.txt}; 
\addplot3[cyan,mark=*,
     x filter/.code={
        \ifnum\coordindex<68
             \def\pgfmathresult{}
         \fi
         \ifnum\coordindex>84
             \def\pgfmathresult{}
         \fi
     },
]
 table [x index = 2,
        y index = 3,
        z index = 4
 ]
{plots/NACA_ReEffect_V1_V2_V3_old.txt}; 
\addplot3[cyan,mark=*,
     x filter/.code={
        \ifnum\coordindex<119
             \def\pgfmathresult{}
         \fi
         \ifnum\coordindex>135
             \def\pgfmathresult{}
         \fi
     },
]
 table [x index = 2,
        y index = 3,
        z index = 4
 ]
{plots/NACA_ReEffect_V1_V2_V3_old.txt}; 
\addplot3[cyan,mark=*,
     x filter/.code={
        \ifnum\coordindex<170
             \def\pgfmathresult{}
         \fi
         \ifnum\coordindex>186
             \def\pgfmathresult{}
         \fi
     },
]
 table [x index = 2,
        y index = 3,
        z index = 4
 ]
{plots/NACA_ReEffect_V1_V2_V3_old.txt}; 

\addplot3[orange,mark=o,
     x filter/.code={
        \ifnum\coordindex<34
             \def\pgfmathresult{}
         \fi
         \ifnum\coordindex>50
             \def\pgfmathresult{}
         \fi
     },
]
 table [x index = 2,
        y index = 3,
        z index = 4
 ]
{plots/NACA_ReEffect_V1_V2_V3_old.txt}; 
\addplot3[orange,mark=o,
     x filter/.code={
        \ifnum\coordindex<85
             \def\pgfmathresult{}
         \fi
         \ifnum\coordindex>101
             \def\pgfmathresult{}
         \fi
     },
]
 table [x index = 2,
        y index = 3,
        z index = 4
 ]
{plots/NACA_ReEffect_V1_V2_V3_old.txt}; 
\addplot3[orange,mark=o,
     x filter/.code={
        \ifnum\coordindex<136
             \def\pgfmathresult{}
         \fi
         \ifnum\coordindex>152
             \def\pgfmathresult{}
         \fi
     },
]
 table [x index = 2,
        y index = 3,
        z index = 4
 ]
{plots/NACA_ReEffect_V1_V2_V3_old.txt}; 
\addplot3[orange,mark=o,
     x filter/.code={
        \ifnum\coordindex<187
             \def\pgfmathresult{}
         \fi
         \ifnum\coordindex>203
             \def\pgfmathresult{}
         \fi
     },
]
 table [x index = 2,
        y index = 3,
        z index = 4
 ]
{plots/NACA_ReEffect_V1_V2_V3_old.txt};

 \addplot3[red,mark=x,
 x filter/.code={
        \ifnum\coordindex<0
             \def\pgfmathresult{}
         \fi
         \ifnum\coordindex>16
             \def\pgfmathresult{}
         \fi
     },
 ]
  table [x index = 2,
        y index = 3,
        z index = 4
 ]
{plots/NACA_Euler_V1_V2_V3_old.txt};
\addplot3[red,mark=x,
 x filter/.code={
        \ifnum\coordindex<17
             \def\pgfmathresult{}
         \fi
         \ifnum\coordindex>33
             \def\pgfmathresult{}
         \fi
     },
 ]
  table [x index = 2,
        y index = 3,
        z index = 4
 ]
{plots/NACA_Euler_V1_V2_V3_old.txt};
\addplot3[red,mark=x,
 x filter/.code={
        \ifnum\coordindex<34
             \def\pgfmathresult{}
         \fi
         \ifnum\coordindex>50
             \def\pgfmathresult{}
         \fi
     },
 ]
  table [x index = 2,
        y index = 3,
        z index = 4
 ]
{plots/NACA_Euler_V1_V2_V3_old.txt};
\addplot3[red,mark=x,
 x filter/.code={
        \ifnum\coordindex<351
             \def\pgfmathresult{}
         \fi
         \ifnum\coordindex>67
             \def\pgfmathresult{}
         \fi
     },
 ]
  table [x index = 2,
        y index = 3,
        z index = 4
 ]
{plots/NACA_Euler_V1_V2_V3_old.txt};

\end{axis}
\end{tikzpicture}
\caption{NACA db 2. DDLS obtained by multi-fidelity AESTAR. 
\textcolor{blue}{$\square$}: $Re_\infty = 10^5$.
\textcolor{cyan}{$\bullet$}: $Re_\infty = 10^6$. 
\textcolor{orange}{$\circ$}: $Re_\infty = 5 \times10^6$. 
\textcolor{red}{$\times$}: $Re_\infty = \infty$.
}
\label{fig.v1v2v3}
\end{figure}

The obtained three-dimensional latent space is shown in figure \ref{fig.v1v2v3}. All airfoils at all $\alpha$ and $Re_\infty$ are displayed with the lines following the AoA.
Much more clear views are obtained looking at the projection on the 
$(V_1,V_2)$ plane (figure \ref{fig.Rev1v2}(a)) and 
$(V_2,V_3)$ plane (figure \ref{fig.Rev1v2}(b)).
It should not surprise that  figure \ref{fig.Rev1v2}(a) is  similar to the DDLSs proposed in section \ref{sec.thin}. Indeed for the case inviscid linear theory the PLS was characterized by the physical  parameters $(\alpha, m)$ whereas here the PLS is given by the physical parameter couple $(\alpha, t)$. The lift versus AoA pictures suggests that the linearity for small AoA increases while increasing thickness; it is reflected in figure \ref{fig.Rev1v2}(a) where iso-$t$ lines become more straight for larger $t$.

The expected correlation of $V_3$ with the Reynolds number is evident in figure \ref{fig.Rev1v2}(b), with a decreasing of $V_3$ while $Re_\infty$ increases.   
\begin{figure}
\centering
\begin{subfigure}{1\linewidth}
\begin{tikzpicture}
\begin{axis}[width=8.0cm,height=8.0cm,
grid,xlabel=$V_1$, ylabel=$V_2$,xmin=-1.0,xmax=1.0,
            ymin=-0.5,ymax=2.5]
\addplot[blue,mark=square,
     x filter/.code={
        \ifnum\coordindex<0
             \def\pgfmathresult{}
         \fi
         \ifnum\coordindex>16
             \def\pgfmathresult{}
         \fi
     },
]
 table [x index = 2,
        y index = 3,
 ]
{plots/NACA_ReEffect_V1_V2_V3_old.txt}; 
\addplot[blue,mark=square,
     x filter/.code={
        \ifnum\coordindex<51
             \def\pgfmathresult{}
         \fi
         \ifnum\coordindex>67
             \def\pgfmathresult{}
         \fi
     },
]
 table [x index = 2,
        y index = 3,
 ]
{plots/NACA_ReEffect_V1_V2_V3_old.txt}; 
\addplot[blue,mark=square,
     x filter/.code={
        \ifnum\coordindex<102
             \def\pgfmathresult{}
         \fi
         \ifnum\coordindex>118
             \def\pgfmathresult{}
         \fi
     },
]
 table [x index = 2,
        y index = 3,
 ]
{plots/NACA_ReEffect_V1_V2_V3_old.txt}; 
\addplot[blue,mark=square,
     x filter/.code={
        \ifnum\coordindex<153
             \def\pgfmathresult{}
         \fi
         \ifnum\coordindex>169
             \def\pgfmathresult{}
         \fi
     },
]
 table [x index = 2,
        y index = 3,
 ]
{plots/NACA_ReEffect_V1_V2_V3_old.txt}; 

\addplot[cyan,mark=*,
     x filter/.code={
        \ifnum\coordindex<17
             \def\pgfmathresult{}
         \fi
         \ifnum\coordindex>33
             \def\pgfmathresult{}
         \fi
     },
]
 table [x index = 2,
        y index = 3,
 ]
{plots/NACA_ReEffect_V1_V2_V3_old.txt}; 
\addplot[cyan,mark=*,
     x filter/.code={
        \ifnum\coordindex<68
             \def\pgfmathresult{}
         \fi
         \ifnum\coordindex>84
             \def\pgfmathresult{}
         \fi
     },
]
 table [x index = 2,
        y index = 3,
 ]
{plots/NACA_ReEffect_V1_V2_V3_old.txt}; 
\addplot[cyan,mark=*,
     x filter/.code={
        \ifnum\coordindex<119
             \def\pgfmathresult{}
         \fi
         \ifnum\coordindex>135
             \def\pgfmathresult{}
         \fi
     },
]
 table [x index = 2,
        y index = 3,
 ]
{plots/NACA_ReEffect_V1_V2_V3_old.txt}; 
\addplot[cyan,mark=*,
     x filter/.code={
        \ifnum\coordindex<170
             \def\pgfmathresult{}
         \fi
         \ifnum\coordindex>186
             \def\pgfmathresult{}
         \fi
     },
]
 table [x index = 2,
        y index = 3,
 ]
{plots/NACA_ReEffect_V1_V2_V3_old.txt}; 

\addplot[orange,mark=o,
     x filter/.code={
        \ifnum\coordindex<34
             \def\pgfmathresult{}
         \fi
         \ifnum\coordindex>50
             \def\pgfmathresult{}
         \fi
     },
]
 table [x index = 2,
        y index = 3,
 ]
{plots/NACA_ReEffect_V1_V2_V3_old.txt}; 
\addplot[orange,mark=o,
     x filter/.code={
        \ifnum\coordindex<85
             \def\pgfmathresult{}
         \fi
         \ifnum\coordindex>101
             \def\pgfmathresult{}
         \fi
     },
]
 table [x index = 2,
        y index = 3,
 ]
{plots/NACA_ReEffect_V1_V2_V3_old.txt}; 
\addplot[orange,mark=o,
     x filter/.code={
        \ifnum\coordindex<136
             \def\pgfmathresult{}
         \fi
         \ifnum\coordindex>152
             \def\pgfmathresult{}
         \fi
     },
]
 table [x index = 2,
        y index = 3,
 ]
{plots/NACA_ReEffect_V1_V2_V3_old.txt}; 
\addplot[orange,mark=o,
     x filter/.code={
        \ifnum\coordindex<187
             \def\pgfmathresult{}
         \fi
         \ifnum\coordindex>203
             \def\pgfmathresult{}
         \fi
     },
]
 table [x index = 2,
        y index = 3,
 ]
{plots/NACA_ReEffect_V1_V2_V3_old.txt};

 \addplot[red,mark=x,
 x filter/.code={
        \ifnum\coordindex<0
             \def\pgfmathresult{}
         \fi
         \ifnum\coordindex>16
             \def\pgfmathresult{}
         \fi
     },
 ]
  table [x index = 2,
        y index = 3,
 ]
{plots/NACA_Euler_V1_V2_V3_old.txt};
\addplot[red,mark=x,
 x filter/.code={
        \ifnum\coordindex<17
             \def\pgfmathresult{}
         \fi
         \ifnum\coordindex>33
             \def\pgfmathresult{}
         \fi
     },
 ]
  table [x index = 2,
        y index = 3,
 ]
{plots/NACA_Euler_V1_V2_V3_old.txt};
\addplot[red,mark=x,
 x filter/.code={
        \ifnum\coordindex<34
             \def\pgfmathresult{}
         \fi
         \ifnum\coordindex>50
             \def\pgfmathresult{}
         \fi
     },
 ]
  table [x index = 2,
        y index = 3,
 ]
{plots/NACA_Euler_V1_V2_V3_old.txt};
\addplot[red,mark=x,
 x filter/.code={
        \ifnum\coordindex<351
             \def\pgfmathresult{}
         \fi
         \ifnum\coordindex>67
             \def\pgfmathresult{}
         \fi
     },
 ]
  table [x index = 2,
        y index = 3,
 ]
{plots/NACA_Euler_V1_V2_V3_old.txt};

 \draw[->, thick]
    (axis cs:-0.68,0.5) --
    (axis cs:-0.5,1)
    node[midway, above, sloped] {$\alpha$};
 \draw[->, thick]
    (axis cs:0.23,1.6) --
    (axis cs:0.5,1.25)
    node[midway, above, sloped] {$t$};
\end{axis}
\end{tikzpicture}
\caption{}
\end{subfigure}

\begin{subfigure}{1\linewidth}
\begin{tikzpicture}
\begin{axis}[width=8cm,height=8.0cm,
grid,xlabel=$V_2$, ylabel=$V_3$,
xmin=-0.5, xmax=2.5,
ymin=-2.5, ymax=0.5]
\addplot[blue,mark=square,
     x filter/.code={
        \ifnum\coordindex<0
             \def\pgfmathresult{}
         \fi
         \ifnum\coordindex>16
             \def\pgfmathresult{}
         \fi
     },
]
 table [x index = 3,
        y index = 4,
 ]
{plots/NACA_ReEffect_V1_V2_V3_old.txt}; 
\addplot[blue,mark=square,
     x filter/.code={
        \ifnum\coordindex<51
             \def\pgfmathresult{}
         \fi
         \ifnum\coordindex>67
             \def\pgfmathresult{}
         \fi
     },
]
 table [x index = 3,
        y index = 4,
 ]
{plots/NACA_ReEffect_V1_V2_V3_old.txt}; 
\addplot[blue,mark=square,
     x filter/.code={
        \ifnum\coordindex<102
             \def\pgfmathresult{}
         \fi
         \ifnum\coordindex>118
             \def\pgfmathresult{}
         \fi
     },
]
 table [x index = 3,
        y index = 4,
 ]
{plots/NACA_ReEffect_V1_V2_V3_old.txt}; 
\addplot[blue,mark=square,
     x filter/.code={
        \ifnum\coordindex<153
             \def\pgfmathresult{}
         \fi
         \ifnum\coordindex>169
             \def\pgfmathresult{}
         \fi
     },
]
 table [x index = 3,
        y index = 4,
 ]
{plots/NACA_ReEffect_V1_V2_V3_old.txt}; 

\addplot[cyan,mark=*,
     x filter/.code={
        \ifnum\coordindex<17
             \def\pgfmathresult{}
         \fi
         \ifnum\coordindex>33
             \def\pgfmathresult{}
         \fi
     },
]
 table [x index = 3,
        y index = 4,
 ]
{plots/NACA_ReEffect_V1_V2_V3_old.txt}; 
\addplot[cyan,mark=*,
     x filter/.code={
        \ifnum\coordindex<68
             \def\pgfmathresult{}
         \fi
         \ifnum\coordindex>84
             \def\pgfmathresult{}
         \fi
     },
]
 table [x index = 3,
        y index = 4,
 ]
{plots/NACA_ReEffect_V1_V2_V3_old.txt}; 
\addplot[cyan,mark=*,
     x filter/.code={
        \ifnum\coordindex<119
             \def\pgfmathresult{}
         \fi
         \ifnum\coordindex>135
             \def\pgfmathresult{}
         \fi
     },
]
 table [x index = 3,
        y index = 4,
 ]
{plots/NACA_ReEffect_V1_V2_V3_old.txt}; 
\addplot[cyan,mark=*,
     x filter/.code={
        \ifnum\coordindex<170
             \def\pgfmathresult{}
         \fi
         \ifnum\coordindex>186
             \def\pgfmathresult{}
         \fi
     },
]
 table [x index = 3,
        y index = 4,
 ]
{plots/NACA_ReEffect_V1_V2_V3_old.txt}; 

\addplot[orange,mark=o,
     x filter/.code={
        \ifnum\coordindex<34
             \def\pgfmathresult{}
         \fi
         \ifnum\coordindex>50
             \def\pgfmathresult{}
         \fi
     },
]
 table [x index = 3,
        y index = 4,
 ]
{plots/NACA_ReEffect_V1_V2_V3_old.txt}; 
\addplot[orange,mark=o,
     x filter/.code={
        \ifnum\coordindex<85
             \def\pgfmathresult{}
         \fi
         \ifnum\coordindex>101
             \def\pgfmathresult{}
         \fi
     },
]
 table [x index = 3,
        y index = 4,
 ]
{plots/NACA_ReEffect_V1_V2_V3_old.txt}; 
\addplot[orange,mark=o,
     x filter/.code={
        \ifnum\coordindex<136
             \def\pgfmathresult{}
         \fi
         \ifnum\coordindex>152
             \def\pgfmathresult{}
         \fi
     },
]
 table [x index = 3,
        y index = 4,
 ]
{plots/NACA_ReEffect_V1_V2_V3_old.txt}; 
\addplot[orange,mark=o,
     x filter/.code={
        \ifnum\coordindex<187
             \def\pgfmathresult{}
         \fi
         \ifnum\coordindex>203
             \def\pgfmathresult{}
         \fi
     },
]
 table [x index = 3,
        y index = 4,
 ]
{plots/NACA_ReEffect_V1_V2_V3_old.txt};

 \addplot[red,mark=x,
 x filter/.code={
        \ifnum\coordindex<0
             \def\pgfmathresult{}
         \fi
         \ifnum\coordindex>16
             \def\pgfmathresult{}
         \fi
     },
 ]
  table [x index = 3,
        y index = 4,
 ]
{plots/NACA_Euler_V1_V2_V3_old.txt};
\addplot[red,mark=x,
 x filter/.code={
        \ifnum\coordindex<17
             \def\pgfmathresult{}
         \fi
         \ifnum\coordindex>33
             \def\pgfmathresult{}
         \fi
     },
 ]
  table [x index = 3,
        y index = 4,
 ]
{plots/NACA_Euler_V1_V2_V3_old.txt};
\addplot[red,mark=x,
 x filter/.code={
        \ifnum\coordindex<34
             \def\pgfmathresult{}
         \fi
         \ifnum\coordindex>50
             \def\pgfmathresult{}
         \fi
     },
 ]
  table [x index = 3,
        y index = 4,
 ]
{plots/NACA_Euler_V1_V2_V3_old.txt};
\addplot[red,mark=x,
 x filter/.code={
        \ifnum\coordindex<351
             \def\pgfmathresult{}
         \fi
         \ifnum\coordindex>67
             \def\pgfmathresult{}
         \fi
     },
 ]
  table [x index = 3,
        y index = 4,
 ]
{plots/NACA_Euler_V1_V2_V3_old.txt};

 \draw[->, thick]
    (axis cs:0.5,0.25) --
    (axis cs:1.1,0.25)
    node[midway, above, sloped] {$\alpha$};
 \draw[->, thick]
    (axis cs:2,0.25) --
    (axis cs:2,-0.4)
    node[midway, above, sloped] {$t$};
\end{axis}
\end{tikzpicture}
\caption{}
\end{subfigure}

\caption{NACA db 2. DDLSs obtained by multi-fidelity AESTAR. 
a): $(V1,V_2)$ plane. b): $(V_2,V_3)$ plane.
\textcolor{blue}{$\square$}: $Re_\infty = 10^5$.
\textcolor{cyan}{$\bullet$}: $Re_\infty = 10^6$. 
\textcolor{orange}{$\circ$}: $Re_\infty = 5 \times10^6$. 
\textcolor{red}{$\times$}: $Re_\infty = \infty$.
}
\label{fig.Rev1v2}
\end{figure}

\begin{figure}
    \centering
   \includegraphics[width=0.6\linewidth]{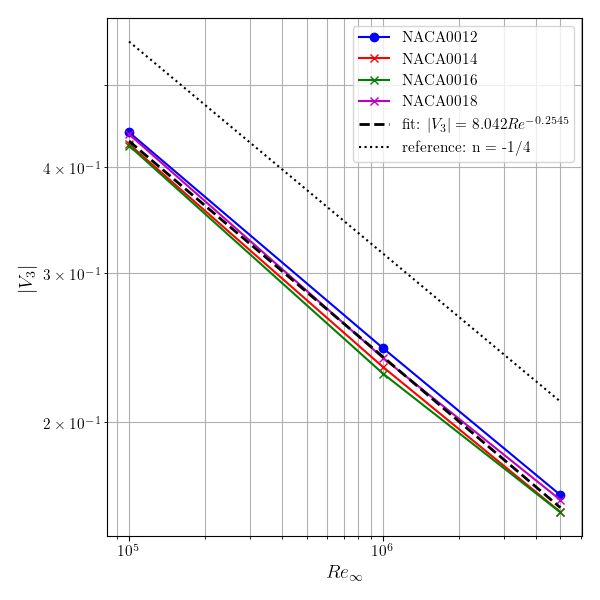}
   \caption{$|V_3|$ at $\alpha=0^\circ$ for all airfoils against Reynolds number and comparison with a log-log power law.}
   \label{fig.V3Re}
\end{figure}

We performed an overall power-law regression using all airfoils data to correlate $V_3$ at $\alpha = 0^\circ$ with the Reynolds number:
\begin{equation}
    |V_3| = A \, Re_\infty^n 
    \label{eq.V3Re}
\end{equation}
with $A = 8.042
\pm 0.5025$, $n = -0.2545\pm 0.004568$ with $R^2 = 0.9968$
and residual standard error in log space $Res= 0.02540$.
The result is nicely evidenced in figure \ref{fig.V3Re}: we obtained a very strict relation between a latent variable and a physical parameter with power exponent $n=-1/4$.
This exponent  characterizes the learned coordinate, whose scale is set by the training (including the weight regularization) and could be not unique. Anyway, the robust findings are that $|V_3|$  varies monotonically with Re, consistently across airfoils, and that the decoder maps it to the correct viscous correction.

The trained decoder is a \emph{multi-fidelity} model.
Indeed, using the first two latent variables and setting $V_3=0$, the decoder returns an inviscid flowfield; while with $V_3\neq0$ it returns a viscous flow: 
{\em the latent variable $V_3$  represents the whole effect of viscosity}.

\begin{figure*}
\centering
\includegraphics[width=0.8\textwidth]{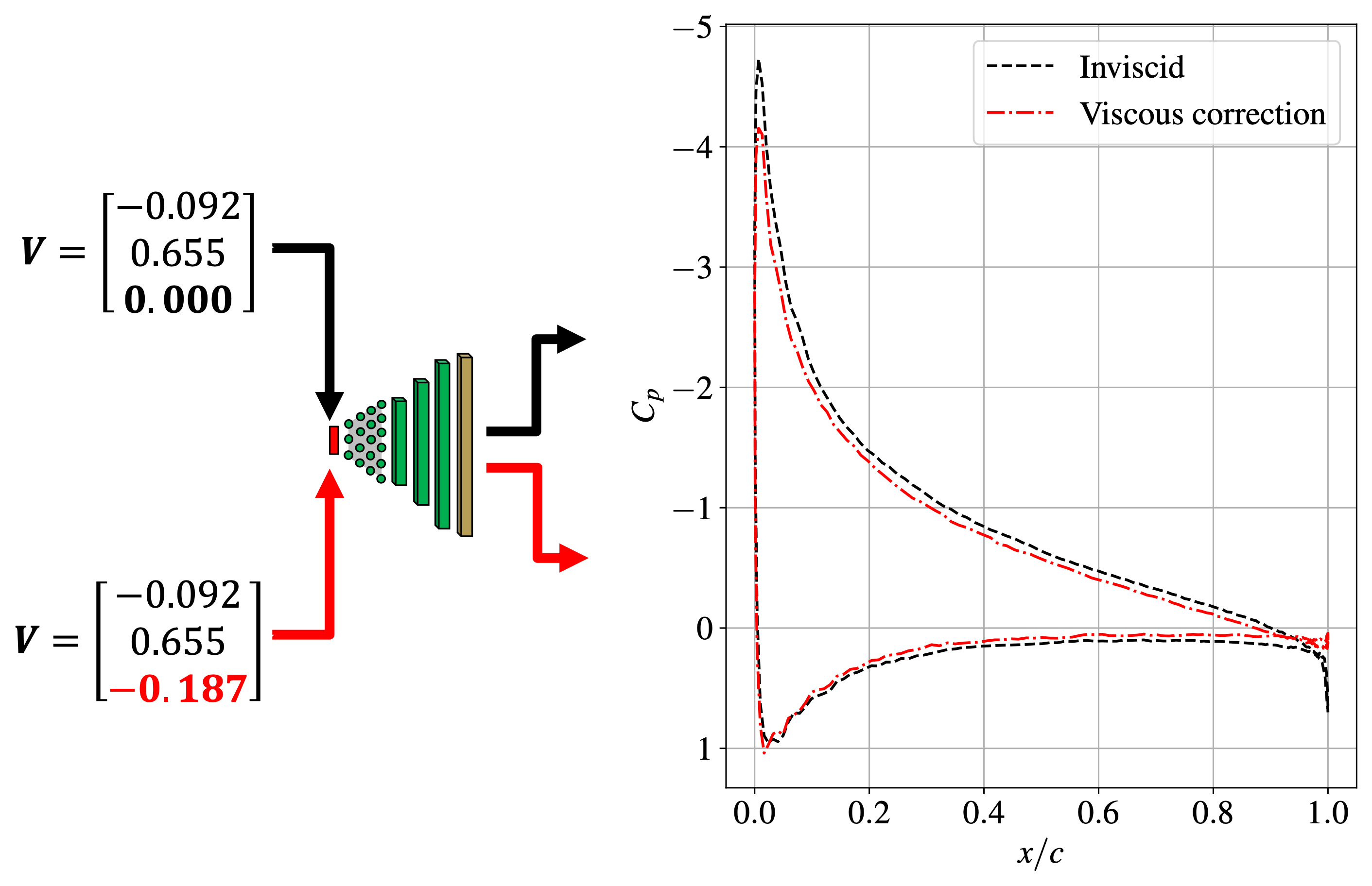}
    \caption{Multi-fidelity decoding. Displayed pressure coefficient on the airfoil surface. NACA 0014 airfoil, $\alpha=10^\circ$, $Re_\infty=5\times10^6$, $M_\infty=0.15$.}
    \label{fig.MF_pred}
\end{figure*}

\begin{figure}
\centering
\includegraphics[width=0.7\linewidth]{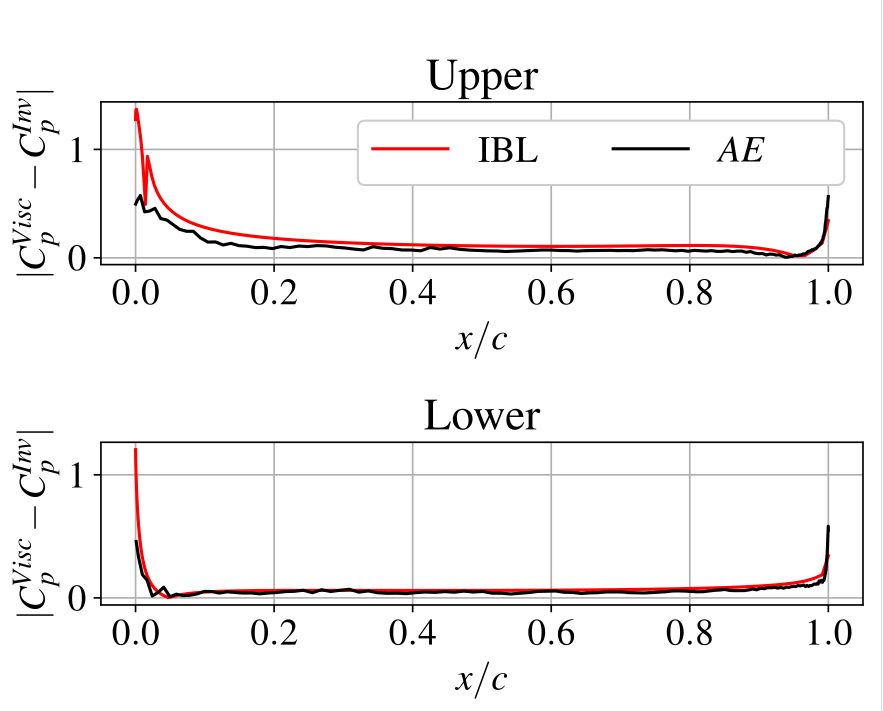}
    \caption{NACA 0014, $\alpha=10^\circ$, $Re_\infty=5\times10^6$, $M_\infty=0.15$. Differences between viscous and inviscid surface pressure distributions by multi-fidelity AESTAR and IBL code. }
    \label{fig.MF_pred_error}
\end{figure}

Figure \ref{fig.MF_pred} shows the prediction of the pressure coefficient around a NACA 0014 airfoil at $\alpha=10^\circ$ obtained by the decoder of the multi-fidelity AESTAR.
In black the inviscid solution is reported, obtained using the corresponding latent vector and  $V_3=0$.  The viscous correction introduced by $V_3$ is shown in red.
It well reproduces the effect of the boundary layer on the pressure distribution.
A quantitative evidence is obtained  comparing present multi-fidelity analysis with a classical Interactive Boundary Layer (IBL) solution of the same airfoil flow obtained by the open source Xfoil software \cite{xfoil}.
Figure \ref{fig.MF_pred_error} shows the surface distribution of the differences between the viscous and inviscid pressure coefficients as predicted by   the multi-fidelity AESTAR and by the IBL code: the agreement is quite convincing. 

While the airfoil lift is essentially related with the AoA,  drag is mainly a function of $Re_\infty$ and airfoil thickness $t$.
The correlation of $V_3$ with drag is shown in figure \ref{fig.V3Cdalpha} where both $|V_3|$ and the drag coefficient $C_d$ are plotted against AoA for the entire database. 
There is a striking equivalence.
 Both $|V_3|$ and $C_d$ grow with AoA and as the Reynolds number decreases.
In addition, $|V_3|$ also accurately detects the drag variation with the thickness of the airfoil,
consider, for instance, the case at constant $Re_\infty=10^5$ (blue curves  in figure \ref{fig.V3Cdalpha}).
\begin{figure*}
\centering

\begin{subfigure}{0.48\textwidth}
\begin{tikzpicture}
\begin{axis}[width=5.7cm,height=8.0cm,
grid,xlabel=$\alpha$($^\circ$), ylabel=$V_3$,
xmin=0, xmax=16,
ymin=-0.001, ymax=2.5]
\addplot[blue,mark=square,
     x filter/.code={
        \ifnum\coordindex<0
             \def\pgfmathresult{}
         \fi
         \ifnum\coordindex>16
             \def\pgfmathresult{}
         \fi
     },
]
 table [x index = 1,
       y expr={-\thisrowno{4}},
 ]
{plots/NACA_ReEffect_V1_V2_V3_old.txt}; 
\addplot[blue,mark=square,
     x filter/.code={
        \ifnum\coordindex<51
             \def\pgfmathresult{}
         \fi
         \ifnum\coordindex>67
             \def\pgfmathresult{}
         \fi
     },
]
 table [x index = 1,
        y expr={-\thisrowno{4}},
 ]
{plots/NACA_ReEffect_V1_V2_V3_old.txt}; 
\addplot[blue,mark=square,
     x filter/.code={
        \ifnum\coordindex<102
             \def\pgfmathresult{}
         \fi
         \ifnum\coordindex>118
             \def\pgfmathresult{}
         \fi
     },
]
 table [x index = 1,
        y expr={-\thisrowno{4}},
 ]
{plots/NACA_ReEffect_V1_V2_V3_old.txt}; 
\addplot[blue,mark=square,
     x filter/.code={
        \ifnum\coordindex<153
             \def\pgfmathresult{}
         \fi
         \ifnum\coordindex>169
             \def\pgfmathresult{}
         \fi
     },
]
 table [x index = 1,
        y expr={-\thisrowno{4}},
 ]
{plots/NACA_ReEffect_V1_V2_V3_old.txt}; 

\addplot[cyan,mark=*,
     x filter/.code={
        \ifnum\coordindex<17
             \def\pgfmathresult{}
         \fi
         \ifnum\coordindex>33
             \def\pgfmathresult{}
         \fi
     },
]
 table [x index = 1,
        y expr={-\thisrowno{4}},
 ]
{plots/NACA_ReEffect_V1_V2_V3_old.txt}; 
\addplot[cyan,mark=*,
     x filter/.code={
        \ifnum\coordindex<68
             \def\pgfmathresult{}
         \fi
         \ifnum\coordindex>84
             \def\pgfmathresult{}
         \fi
     },
]
 table [x index = 1,
        y expr={-\thisrowno{4}},
 ]
{plots/NACA_ReEffect_V1_V2_V3_old.txt}; 
\addplot[cyan,mark=*,
     x filter/.code={
        \ifnum\coordindex<119
             \def\pgfmathresult{}
         \fi
         \ifnum\coordindex>135
             \def\pgfmathresult{}
         \fi
     },
]
 table [x index = 1,
        y expr={-\thisrowno{4}},
 ]
{plots/NACA_ReEffect_V1_V2_V3_old.txt}; 
\addplot[cyan,mark=*,
     x filter/.code={
        \ifnum\coordindex<170
             \def\pgfmathresult{}
         \fi
         \ifnum\coordindex>186
             \def\pgfmathresult{}
         \fi
     },
]
 table [x index = 1,
        y expr={-\thisrowno{4}},
 ]
{plots/NACA_ReEffect_V1_V2_V3_old.txt}; 

\addplot[orange,mark=o,
     x filter/.code={
        \ifnum\coordindex<34
             \def\pgfmathresult{}
         \fi
         \ifnum\coordindex>50
             \def\pgfmathresult{}
         \fi
     },
]
 table [x index = 1,
        y expr={-\thisrowno{4}},
 ]
{plots/NACA_ReEffect_V1_V2_V3_old.txt}; 
\addplot[orange,mark=o,
     x filter/.code={
        \ifnum\coordindex<85
             \def\pgfmathresult{}
         \fi
         \ifnum\coordindex>101
             \def\pgfmathresult{}
         \fi
     },
]
 table [x index = 1,
        y expr={-\thisrowno{4}},
 ]
{plots/NACA_ReEffect_V1_V2_V3_old.txt}; 
\addplot[orange,mark=o,
     x filter/.code={
        \ifnum\coordindex<136
             \def\pgfmathresult{}
         \fi
         \ifnum\coordindex>152
             \def\pgfmathresult{}
         \fi
     },
]
 table [x index = 1,
        y expr={-\thisrowno{4}},
 ]
{plots/NACA_ReEffect_V1_V2_V3_old.txt}; 
\addplot[orange,mark=o,
     x filter/.code={
        \ifnum\coordindex<187
             \def\pgfmathresult{}
         \fi
         \ifnum\coordindex>203
             \def\pgfmathresult{}
         \fi
     },
]
 table [x index = 1,
        y expr={-\thisrowno{4}},
 ]
{plots/NACA_ReEffect_V1_V2_V3_old.txt};

 \addplot[red,mark=x,
 x filter/.code={
        \ifnum\coordindex<0
             \def\pgfmathresult{}
         \fi
         \ifnum\coordindex>16
             \def\pgfmathresult{}
         \fi
     },
 ]
  table [x index = 1,
        y expr={-\thisrowno{4}},
 ]
{plots/NACA_Euler_V1_V2_V3_old.txt};
\addplot[red,mark=x,
 x filter/.code={
        \ifnum\coordindex<17
             \def\pgfmathresult{}
         \fi
         \ifnum\coordindex>33
             \def\pgfmathresult{}
         \fi
     },
 ]
  table [x index = 1,
        y expr={-\thisrowno{4}},
 ]
{plots/NACA_Euler_V1_V2_V3_old.txt};
\addplot[red,mark=x,
 x filter/.code={
        \ifnum\coordindex<34
             \def\pgfmathresult{}
         \fi
         \ifnum\coordindex>50
             \def\pgfmathresult{}
         \fi
     },
 ]
  table [x index = 1,
        y expr={-\thisrowno{4}},
 ]
{plots/NACA_Euler_V1_V2_V3_old.txt};
\addplot[red,mark=x,
 x filter/.code={
        \ifnum\coordindex<351
             \def\pgfmathresult{}
         \fi
         \ifnum\coordindex>67
             \def\pgfmathresult{}
         \fi
     },
 ]
  table [x index = 1,
        y expr={-\thisrowno{4}},
 ]
{plots/NACA_Euler_V1_V2_V3_old.txt};

 \draw[->, thick]
    (axis cs:2,0.3) --
    (axis cs:2,0.8)
    node[midway, above, sloped] {$t$};
\end{axis}
\end{tikzpicture}
\caption{}
\end{subfigure}
\hspace{1mm}
\begin{subfigure}{0.48\textwidth}
\begin{tikzpicture}
\begin{axis}[width=5.7cm,height=8.0cm,
grid,xlabel=$\alpha^\circ$, ylabel=$C_d$,
xmin=0, xmax=16,
ymin=0, ymax=0.25]
\addplot[blue,mark=square]
 table [x expr=\thisrowno{0}, y expr= \thisrowno{2}] 
  {plots/N0012Re.dat}; 
\addplot[blue,mark=square]
 table [x expr=\thisrowno{0}, y expr= \thisrowno{2}] 
  {plots/N0014Re.dat}; 
\addplot[blue,mark=square]
 table [x expr=\thisrowno{0}, y expr= \thisrowno{2}] 
  {plots/N0016Re.dat}; 
\addplot[blue,mark=square]
 table [x expr=\thisrowno{0}, y expr= \thisrowno{2}] 
  {plots/N0018Re.dat};

\addplot[cyan,mark=*]
 table [x expr=\thisrowno{0}, y expr= \thisrowno{4}] 
  {plots/N0012Re.dat}; 
\addplot[cyan,mark=*]
 table [x expr=\thisrowno{0}, y expr= \thisrowno{4}] 
  {plots/N0014Re.dat}; 
\addplot[cyan,mark=*]
 table [x expr=\thisrowno{0}, y expr= \thisrowno{4}] 
  {plots/N0016Re.dat}; 
\addplot[cyan,mark=*]
 table [x expr=\thisrowno{0}, y expr= \thisrowno{4}] 
  {plots/N0018Re.dat}; 

\addplot[orange,mark=o]
 table [x expr=\thisrowno{0}, y expr= \thisrowno{6}] 
  {plots/N0012Re.dat}; 
\addplot[orange,mark=o]
 table [x expr=\thisrowno{0}, y expr= \thisrowno{6}] 
  {plots/N0014Re.dat}; 
\addplot[orange,mark=o]
 table [x expr=\thisrowno{0}, y expr= \thisrowno{6}] 
  {plots/N0016Re.dat};
\addplot[orange,mark=o]
 table [x expr=\thisrowno{0}, y expr= \thisrowno{6}] 
  {plots/N0018Re.dat}; 

\addplot[red,mark=x]
 table [x expr=\thisrowno{0}, y expr= 0] 
  {plots/N0012Re.dat}; 
\addplot[red,mark=x]
 table [x expr=\thisrowno{0}, y expr= 0] 
  {plots/N0014Re.dat};
\addplot[red,mark=x]
 table [x expr=\thisrowno{0}, y expr= 0] 
  {plots/N0016Re.dat}; 
\addplot[red,mark=x]
 table [x expr=\thisrowno{0}, y expr= 0] 
  {plots/N0018Re.dat};

 \draw[->, thick]
    (axis cs:13,0.04) --
    (axis cs:9.5,0.075)
    node[midway, above, sloped] {$t$};
\end{axis}

\end{tikzpicture}
\caption{}
\end{subfigure}

\caption{NACA db 2. DDLSs obtained by multi-fidelity AESTAR. 
a): $V_3$ vs $\alpha$. b) $Cd$ vs $\alpha$.
\textcolor{blue}{$\square$}: $Re_\infty = 10^5$.
\textcolor{cyan}{$\bullet$}: $Re_\infty = 10^6$. 
\textcolor{orange}{$\circ$}: $Re_\infty = 5 \times10^6$. 
\textcolor{red}{$\times$}: $Re_\infty = \infty$.
}
\label{fig.V3Cdalpha}
\end{figure*}
Table \ref{tab.naca2} analyses  the correlation of the obtained three-dimensional latent space with the \emph{physical parameters} $\alpha$ and $t$ and with the \emph{physical responses} lift and drag by ridge regression.
\begin{table}[]
    \centering
    \begin{tabular}{|c|c|c|c|c|}
    \hline
    \hline
    & $\alpha$&$t$&$C_l$&$C_d$\\
  \hline
  \hline
  $V_1,V_2,V_3$ &0.944&0.834&0.903&0.926\\
  \hline
  \hline
    \end{tabular}
    \caption{NACA db 2. Multifidelity AESTAR.
    $R^2$ values of ridge regression
    on physical parameters and responses.
    $R^2$ computed by cross-validation.}
    \label{tab.naca2}
\end{table}
In this case, the correlation of the latent space  is very strong with the physical parameter AoA and with both physical responses lift and drag. Again the latent space is better correlated with the responses of the system    then with  the input parameter thickness.

\begin{figure}
    \centering
    \begin{subfigure}{0.9\linewidth}
    \includegraphics[width=0.6\linewidth]{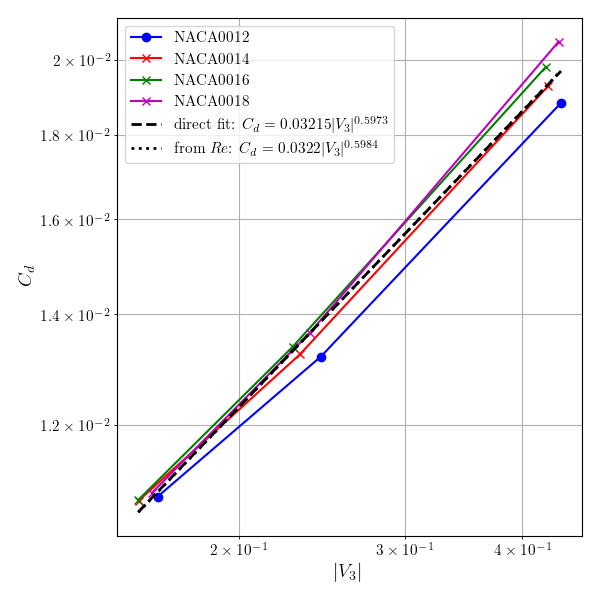}
    \caption{}
    \end{subfigure}
    \begin{subfigure}{0.9\linewidth}
    \includegraphics[width=0.6\linewidth]{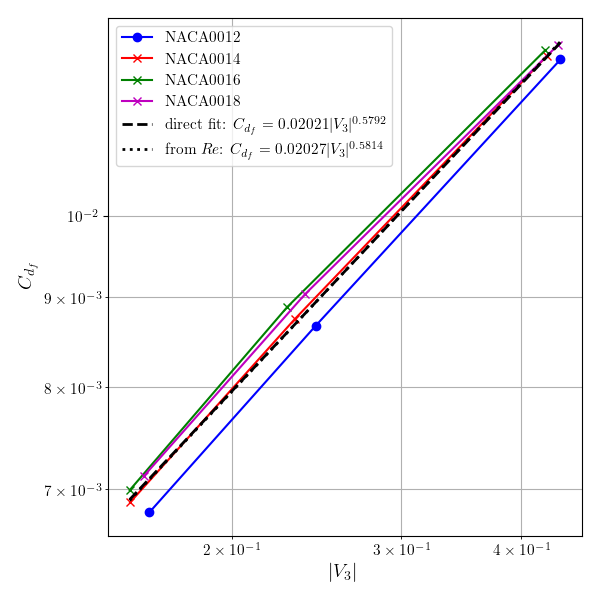}
    \caption{}
    \end{subfigure}
    \caption{NACA db 2. Multifidelity AESTAR. a)  total drag coefficient at $\alpha =0^\circ$ against $|V_3|$. b) friction drag coefficient at $\alpha =0^\circ$ against $|V_3|$.}
    \label{fig.CdV3}
\end{figure}

The obtained log-log power correlation of the drag coefficient at $\alpha=0^\circ$ against $|V_3|$   is:
 \begin{equation}
    C_{d} = B |V_3| ^ m \ ,
    \label{eq.CdV3}
\end{equation}
 where 
 $B=0.03215 \pm 0.0009217$, $m = 0.5973 \pm 0.02001$ with $R^2 = 0.9889$
and residual standard error in log space $Res= 0.02837$. It is displayed in figure
 \ref{fig.CdV3}(a). In the figure it is also reported, for comparison, the $C_d=C_d(|V_3|)$ fit obtained by direct fitting of $C_d$ with $Re_\infty$ and eliminating $Re_\infty$ by eq. \eqref{eq.V3Re}: they are in excellent agreement. It strongly supports the interpretation that the apparent $C_d-V_3$ relationship is not accidental: $V_3$ is encoding the Reynolds-number dependence that also governs drag.

The total drag is built up of the friction contribution essentially depending on the Reynolds number and of the form drag essentially depending on airfoil thickness and AoA. Therefore we can expect a better correlation of $V_3$ with the friction drag coefficient.  the obtained log-log power correlation for this case is:
\begin{equation}
    C_{d_f} = C |V_3| ^ k \ ,
    \label{eq.CdfV3}
\end{equation}
 where 
 $C=0.02021\pm 0.0004827$, $k =  0.5792\pm 0.01667$ with $R^2 = 0.9918$
and residual standard error in log space $Res= 0.02363$. It is proposed in figure \ref{fig.CdV3}(b). The improved fit confirms the better correlation of $V_3$ with friction. The figure also proposes the indirect correlation of $C_{d_f}$
with $V_3$ confirming the agreement.
Eqn. \eqref{eq.V3Re}, \eqref{eq.CdV3} and \eqref{eq.CdfV3}
can be viewed as first examples of quantitative physical scaling relations expressed in terms of latent variables.

\section{Conclusions}

This article examined whether a physically interpretable structure can naturally arise in autoencoders’ latent representations without being explicitly enforced during training. For this purpose, aerodynamic datasets with progressively greater physical complexity were analyzed.
Data-driven latent spaces (DDLSs) obtained by unsupervised vanilla autoencoders were compared with physical latent spaces (PLSs) constructed from \emph{a priori}  known physical parameters and responses.

The results indicate that, even though the latent coordinates are not unique and vary with random initialization, their physical structure is highly consistent: the latent space preserves a \emph{physical imprinting}. In the linear thin-airfoil case, autoencoders trained independently produce latent representations that are nearly perfectly linearly correlated with the physical variables and, with very high accuracy, correspond to affine transformations of the associated PLS. The arbitrariness associated with the orientation, scaling and shear of the latent coordinates therefore does not prevent the underlying physical organization of the solution manifold from being recovered.

The Joukowski problem yields an even more compelling outcome. In the one-dimensional latent setting, the latent variable accurately captures  the minimum pressure coefficient—a true physical response of the system rather than an input parameter used to construct the dataset. When the latent space is two-dimensional, the learned embedding again aligns closely with the physically derived $(\alpha, C_l)$ representation, achieving cross-validated coefficients of determination above 0.99.  In addition, affine transformations of repeated trainings and of the physical response $C_l(\alpha)$ are in remarkably agreement. Overall, these findings suggest that an unsupervised autoencoder can learn to encode not only the parameters that define a physical problem, but also solution-dependent quantities that arise from it.

This observation remains valid when moving to the considerably more complex RANS database including stall, post-stall and massively separated flows. The corresponding DDLS reproduces the ordering and the main nonlinear regimes visible in the aerodynamic responses. Interestingly, the latent representation is more strongly correlated with quantities such as drag rather than with the airfoil thickness used as one of the parameters generating the database. This suggests that the autoencoder does not necessarily organize its representation according to the prescribed input parameters, but may instead preferentially encode the resulting physical response of the system.

The final analysis shows how physical latent structure can also be exploited. Using sequential training, the inviscid solution is represented by the $(V_1,V_2)$ subspace, while the additional variable $V_3$ contains the viscous contribution. Setting $V_3=0$ in the decoder recovers the inviscid field, whereas introducing the appropriate nonzero value of $V_3$ produces the viscous correction, in quantitative agreement with an independent interactive boundary-layer calculation. Moreover, at zero angle of attack the latent variable follows the power law
\begin{equation*}
|V_3|\propto Re_\infty^{-0.2545},
\end{equation*}
remarkably close to an $Re_\infty^{-1/4}$ dependence, and additional power-law relations connect $V_3$ with total and friction drag. The latent coordinate therefore carries quantitative information about the physical mechanism that it represents, rather than acting only as an abstract compression variable.

These observations do not imply that unsupervised autoencoders uniquely identify physical coordinates, nor do they contradict the general non-identifiability of unsupervised latent representations. Instead, they show  that physically meaningful features of the solution manifold can remain reproducible despite the non-uniqueness of the coordinate representation. In this sense, the distinction introduced here between a human-derived PLS and a learned DDLS provides a useful framework for investigating which aspects of a physical system survive data-driven compression.

The present analysis has limitations that will be addressed in future works. (i) Baseline comparisons were limited to PCA in selected cases; additional comparison with linear and nonlinear dimensionality-reduction methods and null models could clarify which features are specific to autoencoders. (ii) The latent organization may depend on data normalization and loss weighting, which determine which flow features dominate the reconstruction error. (iii) Latent coordinates are defined up to invertible transformations, so quantitative relations in latent units are not invariant; physically invariant statements highlight monotonicity and collapse. 

The present results are restricted to aerodynamic databases with deliberately small latent dimensions, and their generality across architectures, physical systems and higher-dimensional manifolds remains to be established. Nevertheless, they suggest that the inspection of highly compressed latent representations may provide more than an explanation of a trained neural network: it may offer a data-driven tool for identifying physical responses, separating physical mechanisms and, potentially, discovering quantitative relations not explicitly supplied during training.

\appendix

\section{Ridge regression and cross-validation}
\label{app.ridge}
Let $\mathbf{x}_i$ and $\mathbf{y}_i$ denote the $i$th  corresponding points of the  PLS and DDLS respectively. For each physical coordinate $x_{ij}$, ridge regression constructs the linear mapping 
\begin{equation} \widehat{x}_{ij} = \beta_{0j}+\mathbf{y}_i^{T}\boldsymbol{\beta}_j , 
\end{equation} where the coefficients are obtained by minimizing 
\begin{equation} \sum_{i=1}^{N} \left(x_{ij}-\beta_{0j} -\mathbf{y}_i^{T}\boldsymbol{\beta}_j\right)^2 +\lambda\left\|\boldsymbol{\beta}_j\right\|_2^2 . 
\end{equation} 
The regularization parameter $\lambda\geq0$ is selected by $K$-fold cross-validation. The predictive correlation between the DDLS and the $j$th physical variable is quantified by 
\begin{equation} R_{\mathrm{CV},j}^{2} = 1- \frac{\displaystyle\sum_{i=1}^{N} \left(x_{ij}-\widehat{x}_{ij}^{\,\mathrm{CV}}\right)^2} {\displaystyle\sum_{i=1}^{N} \left(x_{ij}-\overline{x}_j\right)^2}, 
\end{equation} where $\widehat{x}_{ij}^{\,\mathrm{CV}}$ is the prediction obtained when point $i$ belongs to a validation fold.


\section{Procrustes and affine comparison of latent spaces}
\label{app.procrustes_affine}


\subsection{Orthogonal Procrustes analysis}

Let
$\mathbf{Z}_1,\mathbf{Z}_2 \in \mathbb{R}^{N \times d}$
be two representations of the same $N$ samples in a $d$-dimensional latent space. After centering the two configurations, orthogonal Procrustes analysis determines the orthogonal matrix $\mathbf{R}$ that minimizes
\begin{equation}
\mathbf{R}^{*}
=
\arg\min_{\mathbf{R}^{T}\mathbf{R}=\mathbf{I}}
\left\|
\mathbf{Z}_1\mathbf{R}-\mathbf{Z}_2
\right\|_{F}^{2},
\end{equation}
where $\|\cdot\|_{F}$ denotes the Frobenius norm.

The orthogonality condition
$\mathbf{R}^{T}\mathbf{R}=\mathbf{I}$
restricts the transformation to rotations and reflections. Therefore, distances and angles between points are preserved. If the two latent spaces can be closely aligned by such a transformation, they have essentially the same metric geometry and differ mainly in the arbitrary orientation of their latent coordinates.

The optimal transformation can be obtained from the singular value decomposition
\begin{equation}
\mathbf{Z}_1^{T}\mathbf{Z}_2
=
\mathbf{U}\boldsymbol{\Sigma}\mathbf{V}^{T},
\end{equation}
from which
\begin{equation}
\mathbf{R}^{*}
=
\mathbf{U}\mathbf{V}^{T}.
\end{equation}

A normalized measure of the remaining discrepancy can be defined as
\begin{equation}
E_{P}
=
\frac{
\left\|
\mathbf{Z}_1\mathbf{R}^{*}-\mathbf{Z}_2
\right\|_{F}
}{
\left\|
\mathbf{Z}_2
\right\|_{F}
}.
\end{equation}
A small value of $E_{P}$ indicates that the two representations are nearly equivalent under an orthogonal transformation.

\subsection{General affine transformation}

A less restrictive comparison between the two representations can be obtained using a general affine transformation,
\begin{equation}
\mathbf{Z}_2
=
\mathbf{A}\mathbf{Z}_1+\mathbf{b}\ ,
\end{equation}
where $\mathbf{A}$ is a general nonsingular matrix and $\mathbf{b}$ is a translation vector. After centering the configurations, the translation term is removed and the comparison becomes
\begin{equation}
\mathbf{Z}_2
\simeq
\mathbf{Z}_1\mathbf{A}.
\end{equation}
Unlike an orthogonal transformation, a general affine transformation may include rotation, reflection, unequal scaling along different directions, and shear. Consequently, it preserves straight lines and parallelism, but it does not in general preserve distances, angles, or orthogonality.
For example, an affine transformation can map a rectangular configuration into a parallelogram. In the context of latent representations, such a transformation may indicate that the latent variables learned by an AE are linear combinations of the physically meaningful latent variables.

The optimal affine matrix can be obtained by linear least squares,
\begin{equation}
\mathbf{A}^{*}
=
\arg\min_{\mathbf{A}}
\left\|
\mathbf{Z}_1\mathbf{A}-\mathbf{Z}_2
\right\|_{F}^{2},
\end{equation}
and an analogous normalized residual may be defined as
\begin{equation}
E_{A}
=
\frac{
\left\|
\mathbf{Z}_1\mathbf{A}^{*}-\mathbf{Z}_2
\right\|_{F}
}{
\left\|
\mathbf{Z}_2
\right\|_{F}
}.
\end{equation}

Comparison of $E_{P}$ and $E_{A}$ provides useful information about the relation between two latent spaces. A small Procrustes error indicates that they differ mainly by orientation, whereas a significantly smaller affine error suggests additional linear distortions, such as unequal scaling or shear. If even the affine residual remains large, the relation between the two representations is likely to involve nonlinear distortions.

\bibliography{mlBib}

\end{document}